\documentclass[12pt,a4paper]{article}
\usepackage{jheppub_kim}
 \usepackage{slashed}

\usepackage{longtable}

\usepackage{epsfig}
\usepackage{amssymb,amsmath}
\usepackage{amsthm}
\usepackage{graphicx}
 \usepackage{graphics}
\usepackage[hang,nooneline,scriptsize]{subfigure}
\usepackage{subfigure}
\usepackage{array}

\begin{document}
  
\title{Weak Deflection Angle, Greybody Bound and Shadow for  Charged Massive BTZ Black Hole}
 
\author[a,b] {Sudhaker Upadhyay,\footnote{Visiting Associate at Inter-University Centre for Astronomy and Astrophysics (IUCAA) Pune, Maharashtra-411007.
 }}
 
				\author [c]{Surajit Mandal,}  
				 \author[d]{Yerlan Myrzakulov}
				    \author[d]{ and  Kairat Myrzakulov}
				
	 \affiliation[a]{Department of Physics, K. L. S. College, Magadh University, Nawada, Bihar 805110, India}

 \affiliation[b]{School of Physics, Damghan University, Damghan, 3671641167, Iran}
\affiliation[c]{Department of Physics,   Jadavpur University, Kolkata, West Bengal 700032, India }
	  
  \affiliation[d]{Department of General \& Theoretical Physics, LN Gumilyov Eurasian National University,  Astana, 010008, Kazakhstan}
 
 \emailAdd{sudhakerupadhyay@gmail.com; sudhaker@associates.iucaa.in} \emailAdd{surajitmandalju@gmail.com} 
   \emailAdd{ymyrzakulov@gmail.com}
     \emailAdd{krmyrzakulov@gmail.com }
\abstract{We provide a discussion on a light ray in a charged black hole solution in massive gravity. To serve the purpose, we exploit the optical geometry of the black hole solution and find the 
Gaussian curvature in weak gravitational lensing.  Furthermore, we discuss  the deflection angle of the light ray in both plasma and non-plasma mediums using the Gauss-Bonnet theorem on the black hole.   We also analyze the Regge--Wheeler equation and derive rigorous bounds on the greybody factors of linearly charged massive BTZ black hole. We also study the shadow  or silhouette generated by  charged massive BTZ black holes. The effects of charge  and  cosmological constant   on the radius of the shadow are also discussed. 

}	
	\maketitle 

\section{Introduction}
 Banados, Teitelboim and Zanelli (BTZ) were first to discover three dimensional black holes \cite{btz}. The importance of BTZ black holes lies to the fact that they provide elegant machinery for understanding the lower dimensional gravitational systems and their
interactions \cite{gr}, establishing relation with
string theory \cite{st,st1}, studying various thermal properties of black
holes \cite{th,th1} and many more (see e.g. Refs. \cite{mo,mo1,mo2,mo4,mo5}).
Later,  various three-dimensional black holes along with their thermal properties in different gravity models have been studied \cite{bh1,bh2,bh3,bh4}.
Despite the success of Einstein gravity in low energy limits, there are enough reasons to modify this
theory. Acceleration expansion of the universe and the presence of dark matter and dark
energy are some of these issues. The modification of Einstein gravity by considering massive gravitons solves these problems up to certain extents. In the recent past, 
the BTZ black hole solutions in massive gravity coupled with both the linear and non-linear electrodynamics have been obtained \cite{1}. The details of thermal properties of black holes in various modified gravity
can be found in Refs. 
\cite{01,02,03,04,05,06,07}.

Gravitational lensing is a subject of wide interest that has a tremendous impact on the distribution of matters and
the constituents of the Universe. Gravitational lensing is widely used machinery to explore both populations of both compact 
and extended objects \cite{com, ex}. Weak gravitational lensing has many important aspects in the cosmic microwave background characterization \cite{ch}.  
Within gravitational lensing, the  deflection angle and
the related optical scalars  can be expressed in terms of
derivatives of the independent components of the metrics.
Intriguingly, Gibbons and Werner  proposed a naive elegant
method to  study gravitational
lensing and derived the deflection angle from the Gaussian curvature of the optical
metric \cite{gi}. The theory of weak gravitational lensing in  the generalized gravity is presented in \cite{gi1}. Recently, a weak gravitational lensing of Kerr modified black hole is discussed and found
that  modified gravity effect may appear in gravitational lensing experiments \cite{gi2}.
 
 Greybody factors of black holes are important because they deviate the spectrum
of Hawking radiation of blackbody emission as they are not a perfect blackbody \cite{gre}.   The black-hole greybody
factors can be estimated using various techniques \cite{gre1}.  The greybody factors of the highly rotating black hole signify about the Hawking radiation strong
spin-dependence  \cite{dai}. Greybody Factors of charged dilaton black holes are also 
discussed \cite{sh}. The greybody factors help in calculating the radiation power equation
for the black holes \cite{sud}.
 The low energy
 expression for the greybody factor for the higher-dimensional Schwarzschild \cite{ka} and  black    $dS$ black holes \cite{ka1} coupled with scalar fields have also been discussed.
Recently, the greybody factor  
and   Hawking radiation are estimated for black holes in four-dimensional Einstein-Gauss-Bonnet gravity \cite{ko}. Recently,  the gravitational lensing and greybody bound for the black hole in  Gauss-Bonnet gravity are studied \cite{wah}. Moreover, the greybody factors,   reflection and transmission coefficients are derived for topological
massless black holes in arbitrary dimensions \cite{gon}.  Recently, the  greybody factors and quasinormal modes for various black hole  are reviewed 
in Ref. \cite{sak}. Greybody factors for $d$-dimensional black holes \cite{har},    rotating linear dilaton black holes \cite{sak0},  de Rham-Gabadadze-Tolly black hole in massive gravity \cite{sak01,boo}, non-Abelian charged Lifshitz black branes with $z=2$ hyperscaling violation  \cite{sak02}, Newman-Unti-Tamburino black hole \cite{sak03} and  Schwarzschild-like black hole in the bumblebee gravity \cite{sak1} are also studied.
 This work aims to study the gravitational lensing and bound on greybody factors for the charged BTZ black holes in massive gravity.

On the other hand, due to the strong gravity of the black hole, the two dimensional dark region  occurs  in the celestial sphere called as black hole shadow. 
The concept of the black-hole shadow  appears when  there exists a geometrically
thick and optically thin emission region around the  event horizon of black hole \cite{28}.  It was studied first for the Schwarzschild
black hole \cite{29}. Recently, the weak gravitational lensing and   shadow cast of  generalized Einstein-Cartan-Kibble-Sciama gravity theory are studied \cite{30}.  The shadow cast   generated by a Kerr-Newman-Kasuya   black hole
is discussed in Ref. \cite{31}. 
 
This paper is presented in nine parts. In section \ref{sec3}, we outline a charged black hole solution in massive gravity and demonstrate corresponding optical metric and Gaussian curvature in weak gravitational lensing. In section \ref{sec4}, using the Gauss-Bonnet theorem, we evaluate the 
deflection angle in weak limit for such a black hole in a non-plasma medium.  By doing graphical analysis,   the effects of various parameters on the deflection angle in a non-plasma medium are studied in section \ref{sec40}.   In section \ref{sec5}, within the context of gravitational lensing, we derive  Gaussian optical curvature and, therefore, deflection angle for the considered black hole in the plasma medium. The deflection angle in the plasma medium has additional terms corresponding to the refractive index of the plasma medium. Similar to the non-plasma medium case, the graphical analyses to study the effects of several parameters on deflection angle in plasma medium are also 
presented in section \ref{sec6}.
The rigorous bound on greybody factor (transmission probability) for the linearly charged massive BTZ black hole is estimated in section \ref{sec7}. The behaviors of the potential and bound on greybody factor are given 
in section \ref{sec8}. In section \ref{sec09}, we present discussions related to black hole shadow. The shape of the silhouette of the shadow is 
estimated from  the geodesic equations
of a test particle around the black hole.  
Finally, we conclude the results and make final remarks in the
section \ref{sec9}.

\section{  Linearly charged BTZ black hole in massive gravity }\label{sec3}
In this section,
 we study the  linearly  charged BTZ black hole solution in the context of massive gravity and calculate Gaussian optical curvature for the model
in the framework of weak gravitation lensing.  The  massive BTZ gravity associated with electrodynamics   in $3$-dimensions is described by   following action:
\begin{eqnarray}
{\cal I}=-\frac{1}{16\pi}\int d^3x \sqrt{-g}\left[R-2\Lambda-F_{\mu\nu}F^{\mu\nu}+m^2\sum_i^4c_iU_i(g,f)\right],\label{act}
\end{eqnarray}
where $R$ is a Ricci scalar, $\Lambda$ is a cosmological constant, $F_{\mu\nu}$ is the Maxwell field strength tensor, $m$ represents the graviton mass, and $f$ refers to a fixed symmetric tensor.
 Here, $c_i$ are some constants and $U_i$ are symmetric
polynomials  \cite{1}.
The field equations corresponding to the above action (\ref{act}) are given by \cite{hen}:
\begin{eqnarray}
R_{\mu\nu}-\frac{1}{2}g_{\mu\nu}R +\Lambda g_{\mu\nu}-\frac{1}{2}g_{\mu\nu}
F_{\eta\rho}F^{\eta\rho}+2F_{\mu\eta}F^\eta_\nu +m^2\xi_{\mu\nu}& =&0,\\
\partial_\mu(\sqrt{-g}F^{\mu\nu})& =&0,
\end{eqnarray}
where
\begin{eqnarray}
\xi_{\mu\nu}&=& -\frac{c_1}{2}(U_1g_{\mu\nu}-K_{\mu\nu}) -\frac{c_2}{2}(U_2g_{\mu\nu}-2U_1K_{\mu\nu}+2K_{\mu\nu}^2)-\frac{c_3}{2}(U_3g_{\mu\nu} -3U_2K_{\mu\nu} \nonumber\\
&+&6U_1K_{\mu\nu}^2 -6K_{\mu\nu}^3)-\frac{c_4}{2}(U_4g_{\mu\nu} -4U_3K_{\mu\nu} +12U_2K_{\mu\nu}^2 -24U_1K_{\mu\nu}^3+24K_{\mu\nu}^4).
\end{eqnarray} 
 Here, $K_{\mu\nu}$ is the $d\times d$ matrix defined as $K_{\mu\nu}=\sqrt{g_{\mu\alpha}g^{\alpha\rho}f_{\rho\nu}}$ \cite{1}. 

The black hole solution for the linearly charged BTZ massive gravity is given by 
\begin{equation}\label{1}
  ds^2=-f(r)dt^2+\frac{dr^2}{f(r)}+r^2d\phi^2,
\end{equation}
where metric function takes the following form: 
\begin{equation}\label{2}
f(r)=-\Lambda r^2-m_0-2q^2\ln \left(\frac{r}{l}\right)+m^2cc_1r.
 \end{equation}
Here,  $m_0$  and $q$  are integration constants related to the mass  ($M= m_0/8$) and the electric charge of the black hole ($Q=q/2$), respectively. However, $l$ is an arbitrary length parameter and  $c$ is the positive constant.

Here, we should note that the Reissner-Nordstr\"om   solution describes a charged black hole in asymptotically flat space which corroborates with charged BTZ black hole solution. Meanwhile, a strong gravitational lensing is discussed for the  Reissner-Nordstr\"om  black hole in Ref. \cite{rn}.
The greybody factor of nonminimally coupled scalar fields from Reissner-Nordstr\"om  black hole is presented   in low frequency
approximation \cite{rn1}.
The present analysis of weak gravitational lensing and greybody bound for the  charged massive BTZ black hole  will be totally different than the case of Reissner-Nordstr\"om black hole    studied  in Ref. \cite{rn} and  \cite{rn1} because the solution 
(\ref{2}) does not coincide with the  Reissner-Nordstr\"om  black hole in massless limit.
\subsection{ Optical metric and its Gaussian curvature in weak gravitational lensing}
We now focus on null geodesics deflected by this black hole. 
It is well-known that light satisfies the null geodesic  (i.e. $ds^2=0$). This null geodesic helps in defining the optical metric that describes Riemannian geometry followed by the light. 
Now, corresponding to the null condition,  we have the following optical metric:
 \begin{equation}\label{4}
 dt^2=\bar{g}_{ij} dx^i dx^j= d\tilde{r}^2+f(\tilde{r})^2d\phi^2
   \end{equation}
 where  \begin{equation}\label{6}
 d\tilde{r}=\frac{dr}{(-\Lambda r^2-m_0-2q^2\ln\left(\frac{r}{l}\right)+m^2cc_1r)},
 \end{equation}
 \begin{equation}\label{7}
 f(\tilde{r})=\frac{r}{\sqrt{-\Lambda r^2-m_0-2q^2\ln\left(\frac{r}{l}\right)+m^2cc_1r}}.
 \end{equation}
Now, it is obvious that the equatorial
plane in the optical metric is a surface of revolution.
The non-vanishing Christoffel symbols associated with metric (\ref{4})  are computed as 
\begin{eqnarray}
\Gamma^{\tilde{r}}_{\phi\phi}&=&\frac{r(rf^\prime(\tilde{r})-2f(\tilde{r}))}{2},\\
\Gamma^{\phi}_{\tilde{r}\phi}&=&\frac{2f(\tilde{r})-rf^\prime(\tilde{r})}{2},\\\Gamma^{\tilde{}r}_{\tilde{r}\tilde{r}}&=&-\frac{f^\prime(\tilde{r})}{f(\tilde{r})}.
\end{eqnarray} 
Here, prime denotes derivative with respect to $r$. With the help of above Christoffel symbols, we only have the following non-zero Riemann tensor for optical curvature: $R_{\tilde{r}\phi\tilde{r}\phi}$=-$kf^2(\tilde{r})$.
The Gaussian optical curvature $\mathcal{K}$ is related to the Ricci scalar as
   \begin{equation}\label{8}
   \mathcal{K}=\frac{R}{2}=-\frac{1}{f(\tilde{r})}\left[\frac{d r}{d \tilde{r}} \frac{d}{d r}\left(\frac{d r}{d \tilde{r}}\right) \frac{d f}{d r}+\frac{d^{2} f}{d r^{2}}\left(\frac{d r}{d \tilde{r}}\right)^{2}\right].
   \end{equation}
 Corresponding to equations (\ref{6}) and (\ref{7}), the Gaussian optical curvature  eventually takes the following  explicit form:
\begin{eqnarray}\label{10}
 \mathcal{K}&=&\Lambda m_0-3\Lambda q^2+2\Lambda q^2\ln\left(\frac{r}{l}\right)-\frac{q^2m_0}{r^2}+\frac{6m^2q^2cc_1}{r}\ln\left(\frac{r}{l}\right)+\frac{2m^2q^2cc_1}{r}\nonumber\\
 &+& O(q^4,m^4,c^2,c_1^2).
\end{eqnarray}
Here, it is worth mention  that this  Gaussian optical curvature leads to real valued deflection angle only for negative cosmological constant.
Therefore, we need to consider the negative cosmological constant for further analyses.
To do so,  we replace cosmological constant with its negative value as $ \Lambda^\prime=- \Lambda$ in the Gaussian optical curvature. This  leads to 
 \begin{eqnarray}\label{11}
 \mathcal{K}&=& -\Lambda^\prime m_0+3\Lambda^\prime q^2-2\Lambda^\prime q^2\ln\left(\frac{r}{l}\right)-\frac{q^2m_0}{r^2}+\frac{6m^2q^2cc_1}{r}\ln\left(\frac{r}{l}\right)+\frac{2m^2q^2cc_1}{r}\nonumber\\
 &+&O(q^4,m^4,c^2,c_1^2).
\end{eqnarray}
 Here, one can see that the Gaussian optical curvature depends on various parameters like charge, mass, cosmological constant and length parameter.
 
\section{Deflection angle of charged massive BTZ black hole in non-plasma  medium}\label{sec4}
 In this section, using the Gauss-
 Bonnet theorem, we calculate the deflection angle of a linearly charged 
 massive BTZ  black hole   in the    non-plasma medium.  The Gauss-Bonnet theorem,  which provides a connection  between  the (intrinsic) geometry of metric and its topology in the regular 
 domain $\mathcal{V}_{\cal {R}}$ with boundary $\partial \mathcal{V}_{\cal 
 {R}}$, is expressed as
 \begin{equation}\label{12}
 \iint_{\mathcal{V}_{\cal R}} \mathcal{K} d S+\oint_{\partial \mathcal{V}
 _{\cal R}} k d t+\sum_{z} \alpha_{z}=2 \pi  {\Xi}\left(\mathcal{V}_{\cal 
 {R}}\right),
 \end{equation}
 where  $\mathcal{V}_{\cal R}\subset S$ is a regular domain of  
two-dimensional surface   $S$ with  simple, closed, regular,
piecewise,  and positive oriented boundary $\partial \mathcal{V}_{\cal R}$.  
 Here,  $k$ is the geodesic curvature of $\partial \mathcal{V}_{\cal R}$ 
 given as $k=\bar{g}\left(\nabla_{\dot{\gamma}} \dot{\gamma}, \ddot{\gamma}
 \right)$, where $\gamma$ is a smooth curve 
 of unit speed in such a way $\bar{g}(\dot{\gamma}, \dot{\gamma})=1$ and  $
 \ddot{\gamma}$ is unit acceleration vector.  $\alpha_{z}$ refers to the 
 exterior angle at the $z^{\mbox{th}}$ vertex. Here, $\Xi$ is an Euler 
 characteristic number. 
 In the limit of the radius  ${\cal R} \rightarrow \infty$ (of the curve $E_{\cal R}$),  jump angle takes value $\pi/2$   and 
the characteristic number becomes a unit. 
In this limit, 
 the geodesic curvature can be expressed as $k\left(E_{\cal R}\right)= |\nabla_{\dot{E}_{\cal R}} \dot{E}_{\cal R}|$.  The radial part of geodesic curvature can be written as
\begin{equation}\label{14}
\left(\nabla_{\dot{E}_{\cal R}} \dot{E}_{\cal R}\right)^{r}=\dot{E}_{\cal R}^{\phi} \partial_{\varphi} \dot{E}_{\cal R}^{r}+\Gamma_{\varphi \varphi}^{\tilde{r}}\left(\dot{E}_{\cal R}^{\phi}\right)^{2}.
\end{equation}
For very large ${\cal R}$, the curve  $E_{\cal R}$ is defined by $r(\phi)={\cal R}=$ constant and this leads  to $\left(\dot{E}_{R}^{\phi}\right)^{2}=\frac{1}{f^{2}(\tilde{r})}$.
Corresponding to the Christoffel symbols, in connection to the optical geometry, by memorizing $\Gamma_{\phi \phi}^{\tilde{r}}=\frac{r\left(r f^{\prime}(\tilde{r})-2 f(\tilde{r})\right)}{2}$, we calculated geodesic curvature as 
 \begin{equation}\label{15}
 \left(\nabla_{\dot{E}_{\cal R}^{r}} \dot{E}_{{\cal R}}^{r}\right)^{r} \rightarrow \frac{1}{\cal R}.
 \end{equation}
This implies that $k(E_{\cal R})\rightarrow \frac{1}{\cal R}$. Using optical metric (\ref{4}), we can have $d t={\cal R} d \phi$. Consequently, 
 \begin{equation}\label{16}
 k(E_{\cal R})dt=\lim_{{\cal R}\to\infty}[k(E_{\cal R})dt]
         =\lim_{{\cal R}\to\infty}\left[\frac{1}{2\sqrt{\bar{g}_{rr}  \bar{g}_{\phi\phi}}}(\frac{\partial\bar{g}_{\phi\phi}}{\partial r})\right]d\phi
         =d\phi.
 \end{equation}
 Taking all the discussions  into account, the  Gauss-Bonnet theorem becomes 
 \begin{equation}\label{17}
 \iint_{\mathcal{V}_{{\cal R}}} \mathcal{K} d S+\oint_{\partial \mathcal{V}_{{\cal R}}} k d t\stackrel{{\cal R}\rightarrow \infty} {=}  \iint_{S_{\infty}} \mathcal{K} d S+\int_{0}^{\pi+\tilde{\delta}} d \phi.
 \end{equation}
  
In the weak deflection limit, the light ray  at   the zeroth order follows a straight line approximation as $r(t)=\frac{b}{\sin\varphi}$, where $b$ is the impact parameter. With this, the deflection angle can be written as
 \begin{equation}\label{18}
 \tilde{\delta}=-\int_{0}^{\pi} \int_{b / \sin \phi}^{\infty} \mathcal{K} dS=-\int_{0}^{\pi} \int_{b / \sin \phi}^{\infty} \mathcal{K} \sqrt{\operatorname{det} \bar{g}} d \tilde{r} d \phi =-\int_{0}^{\pi} \int_{b / \sin \phi}^{\infty} \mathcal{K} \frac{r}{f(r)^\frac{3}{2}} d \tilde{r} d \phi.
 \end{equation}
 For the given metric function of linearly charged massive BTZ black hole (\ref{2}) and Gaussian optical curvature (\ref{11}),   the deflection angle   for non-plasma medium simplifies to
   \begin{eqnarray}\label{22}
 \tilde{\delta}&=&-\int_{0}^{\pi} \int_{b / \sin \phi}^{\infty}\left[-\frac{(m_0-3q^2)}{{\Lambda^\prime}^\frac{1}{2}r^2}-\frac{m_0}{{\Lambda^\prime}^\frac{3}{2}r^4}\left(\frac{3m_0}{2}-8q^2\right)+\frac{m^2cc_1}{2{\Lambda^\prime}^\frac{3}{2}r^3}(3m_0-5q^2)\right.
\nonumber\\
&+&\left. O(q^4,m^4,c^2,c_1^2) \right]d \tilde{r} d \phi\nonumber\\
&=&\frac{(m_0-3q^2)}{{\Lambda^\prime}^\frac{1}{2}b}\int^\pi_0 
\sin\phi d\phi +\frac{m_0}{3{\Lambda^\prime}^\frac{3}{2}b^3} \left(\frac{3m_0}{2}-8q^2\right) \int^\pi_0 
\sin^3\phi d\phi \nonumber\\
&-&\frac{m^2cc_1(3m_0-5q^2)}{4{\Lambda^\prime}^\frac{3}{2}b^2}
 \int^\pi_0 
\sin^2\phi d\phi.
 \end{eqnarray}
This, therefore, in the weak limit, gives the explicit expression for the deflection angle for linearly charged massive BTZ black hole as
 \begin{equation}\label{23}
  \tilde{\delta}=\frac{2(m_0-3q^2)}{{\Lambda^\prime}^\frac{1}{2}b}+\frac{4m_0}{9{\Lambda^\prime}^\frac{3}{2}b^3}\left(\frac{3m_0}{2}-8q^2\right)-\frac{m^2cc_1\pi}{8{\Lambda^\prime}^\frac{3}{2}b^2}(3m_0-5q^2)+O(q^4,m^4,c^2,c_1^2).
 \end{equation}
 Here, it is evident that the deflection angle of charged massive BTZ black hole depends on the various parameters like impact parameter $b$,   charge $q$, the mass parameter $m_0$, and cosmological constant.
 
  \section{Graphical analysis for non-plasma medium}\label{sec40}
 In this section, we study the behavior of deflection angle and their dependence on 
 various parameters. 
 \subsection{Effect of impact parameter ($b$) on deflection angle ($ \tilde{\delta}$) }
\begin{figure}[h!]
\begin{center} 
 $\begin{array}{cccc}
\subfigure[]{\includegraphics[width=0.5\linewidth]{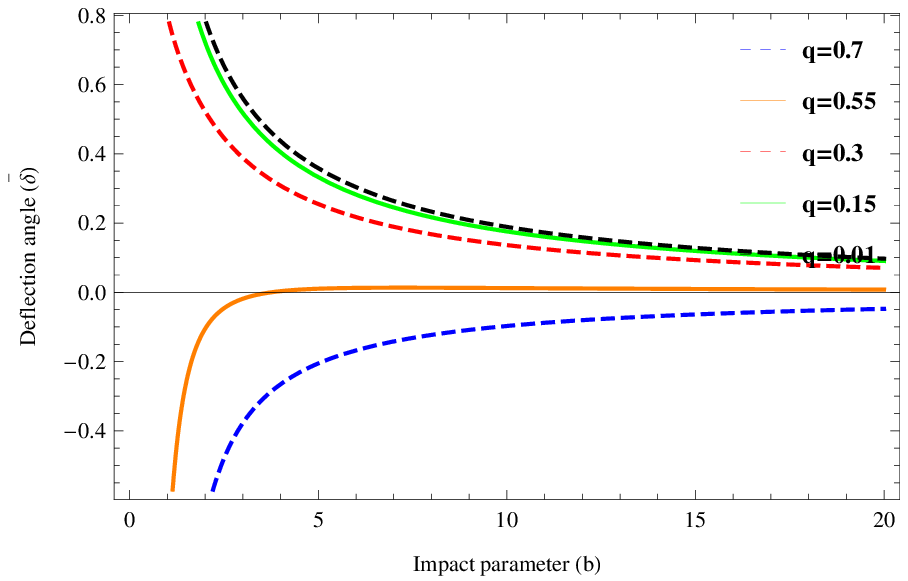}
\label{1a}}
\subfigure[]{\includegraphics[width=0.5\linewidth]{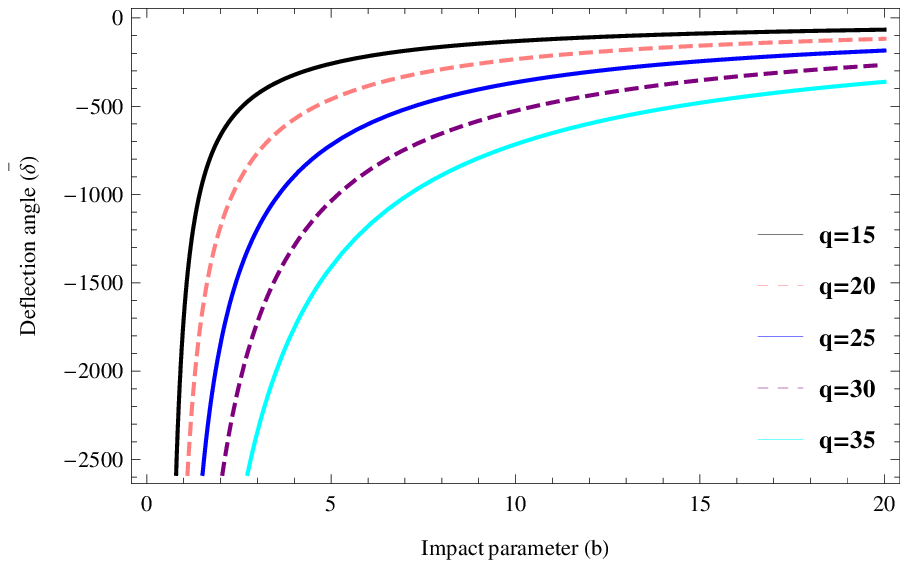}\label{1b}} \\
\subfigure[]{\includegraphics[width=0.5\linewidth]{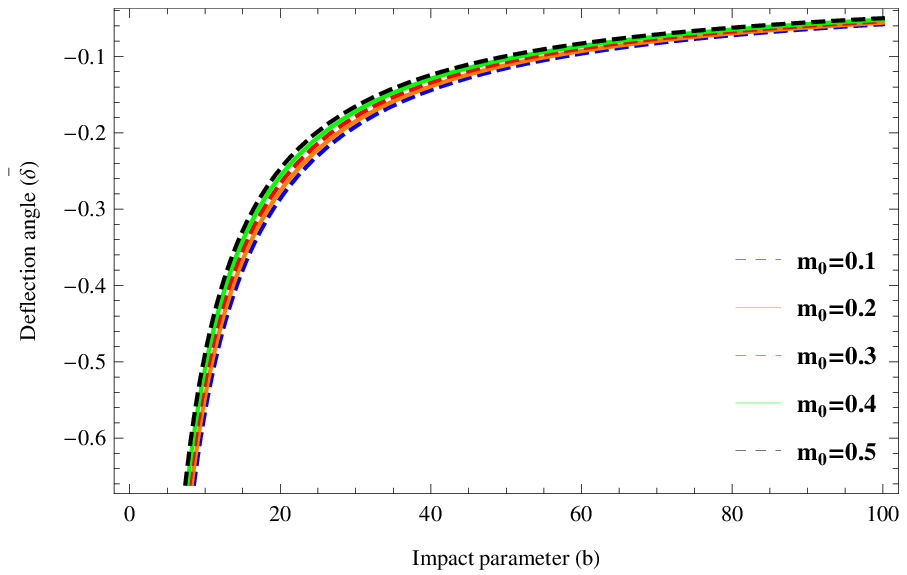}\label{1c}}
\subfigure[]{\includegraphics[width=0.5\linewidth]{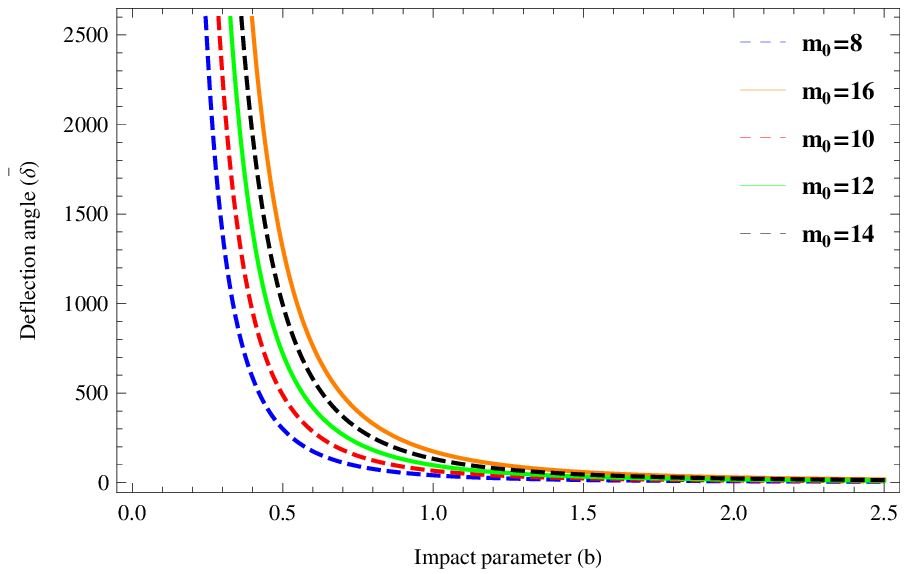} \label{1d}}
\end{array}$
\end{center}
\caption{  In \ref{1a} and \ref{1b}, the behavior of deflection angle ($\tilde{\delta}$) with respect to impact parameter ($b$) for varying $q$ but fixed $m_0=1$. 
 In \ref{1c} and \ref{1d}, the behavior of  $\tilde{\delta}$ with respect to $b$ for varying $m_0$ but fixed $q=1$. Here, $\Lambda'=c=c_1=1$. }
\label{fig1}
\end{figure}

 To discuss the effect of impact parameter on deflection angle, we plot figure \ref{fig1}.
Here, from the plots \ref{1a} and \ref{1b}, it is clear that the deflection angle 
decreases with the impact parameter for very small $q$ but remains positive. In contrast,
for large $q$, the deflection angle increases with $b$ but takes negative values only.
However, from figures \ref{1c} and \ref{1d}, we see that the deflection angle increases with the impact parameter for small $m_0$ and remains negative valued. For larger black hole mass, the deflection angle is asymptotically decreasing function but remains positively valued.

\subsection{Effect of  charge ($q$) on deflection angle ($ \tilde{\delta}$) }
\begin{figure}[ht]
\begin{center}
\includegraphics[width=0.6\linewidth]{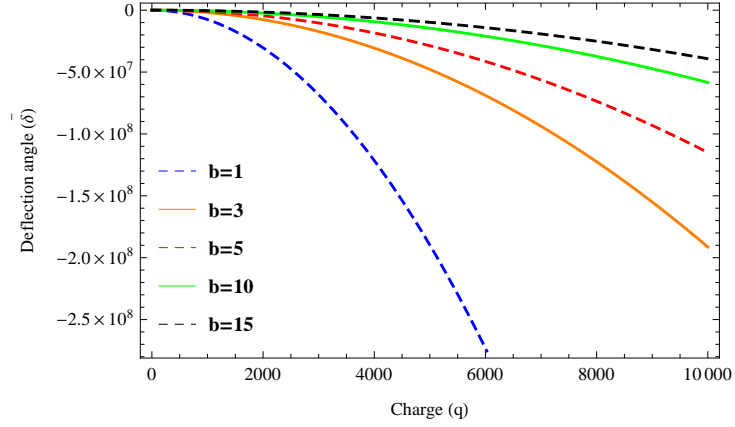} 
\end{center}
\caption{The behavior of deflection angle ($\tilde{\delta}$) with respect to   charge $q$  by varying  $b$. Here, $m=m_0=\Lambda'=c=c_1=1$.}
\label{fig2}
\end{figure}

To study the effect of electric charge $q$  on deflection angle, we plot figure \ref{fig2}.
From the plot, we observe that the deflection angle is a decreasing function of $q$.
The value of deflection angle becomes more negative when impact parameter increases. 

\subsection{Effect of mass parameter ($m$) on deflection angle ($ \tilde{\delta}$) }
\begin{figure}[ht]
  \begin{center} 
  $\begin{array}{cccc}
\subfigure[]{\includegraphics[width=0.5\linewidth]{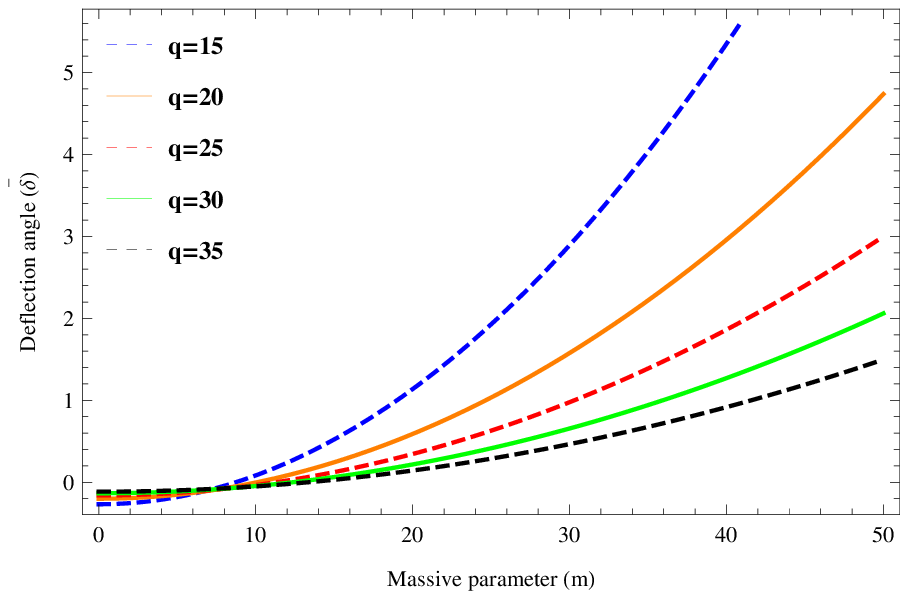}
\label{3a}}
\subfigure[]{\includegraphics[width=0.5\linewidth]{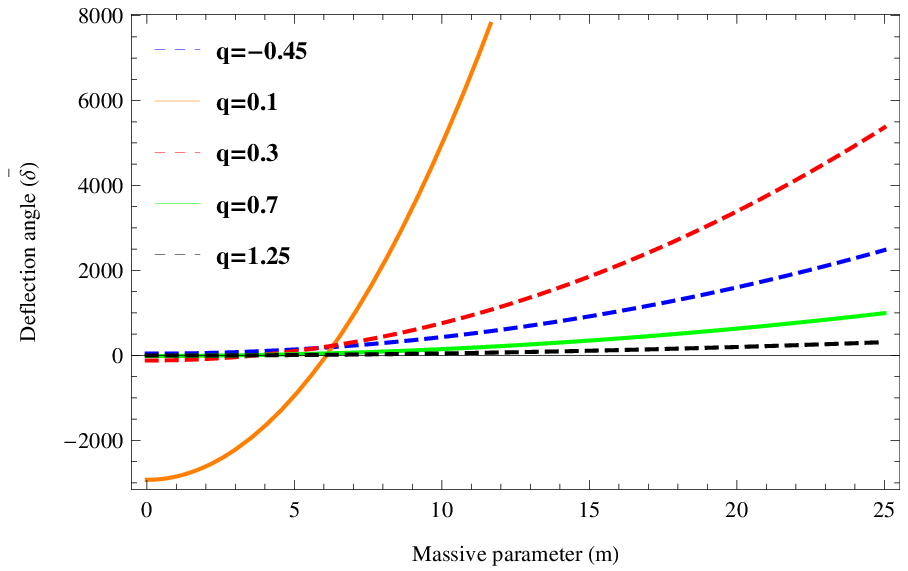}\label{3b}}  
\end{array}$
 \end{center}
\caption{The behavior of $\tilde{\delta}$ with respect to $m$  by changing charge $q$.
Here, $b=m_0=\Lambda'=c=c_1=1$.
 }
\label{fig3}
\end{figure}

To study the effect of the mass parameter ($m$)   on deflection angle, we plot figure \ref{fig3}.
The plot tells that 
the deflection parameter is an increasing function of the mass parameter. There is a critical value for deflection angle that does not depend on the value of $q$.
However, for the larger value of $q$, the deflection angle for massive 
black holes  decreases. In \ref{3b}, we see that for very small $q$, the deflection angle takes a negative value for small $m$. 

 \subsection{Effect of cosmological constant ($\Lambda'$)  on deflection angle ($ \tilde{\delta}$)   } 
\begin{figure}[ht]
\begin{center}  $\begin{array}{cccc}
\subfigure[]{\includegraphics[width=0.5\linewidth]{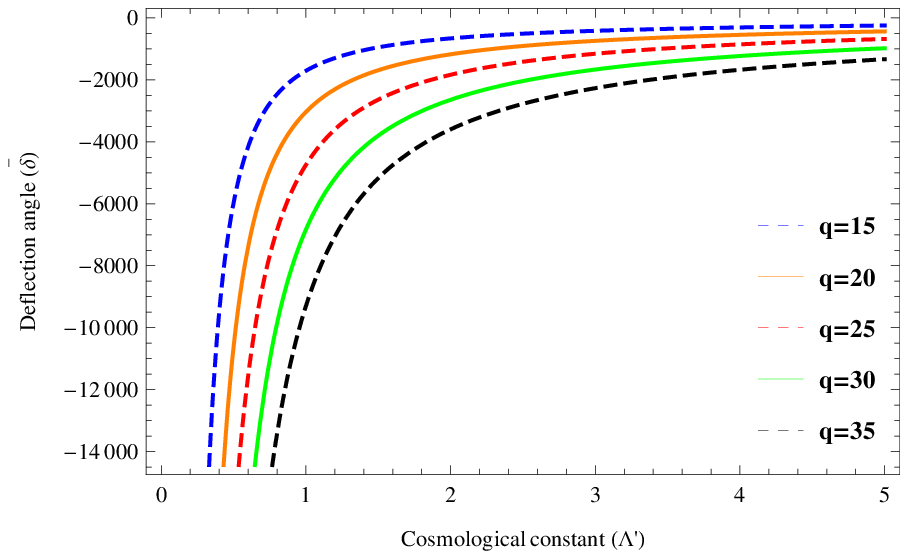}
\label{4a}}
\subfigure[]{\includegraphics[width=0.5\linewidth]{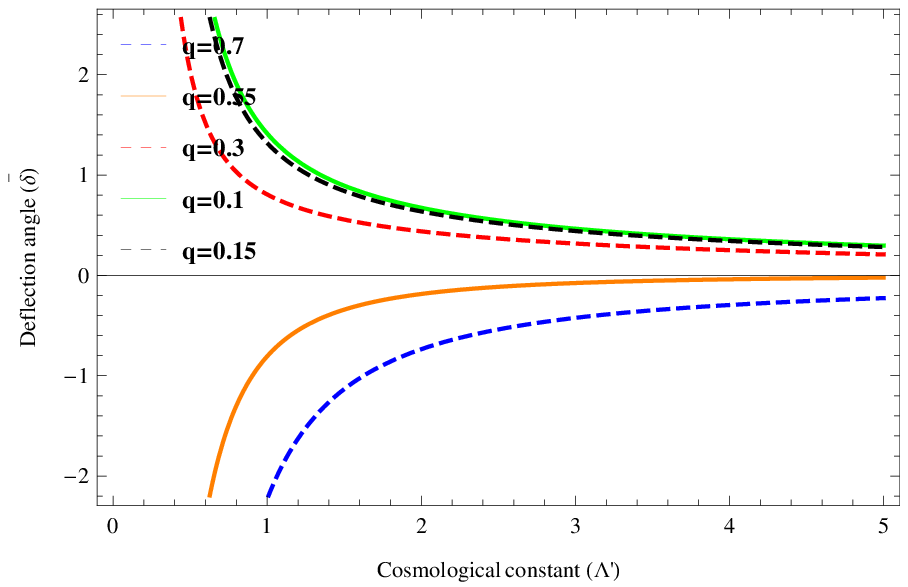}\label{4b}}  
\end{array}$
\end{center}
\caption{ Plot of $\tilde{\delta}$ with respect to $\Lambda'$  by changing $q$.  Here, $b=m_0=m=c=c_1=1$.  
 }
\label{fig4}
\end{figure}

 The impacts of cosmological constant ($\Lambda'$)  on deflection angle ($ \tilde{\delta}$)   are depicted in figure \ref{fig4}. 
 From the figure \ref{4a} and \ref{4b}, we see that deflection angle is both increasing and decreasing function with respect to $\Lambda'$  according  to 
different  $q$. For large $q$, it is negative valued but increases with $q$.
However, for very small values of $q$ it  shows opposite character. 
 \section{Gravitational lensing by linearly charged massive BTZ black hole in plasma medium}\label{sec5}
 In this section, we study   the gravitational lensing of linearly charged massive BTZ black hole  filled with
a cold non-magnetized plasma with the refractive
index $n$. The refractive index  satisfies the following relation \cite{vol}: 
 \begin{equation}\label{24}
 n^{2}(r)=1-\frac{\omega_{e}^{2}(r)}{\omega_{\infty}^{2}(r)}f(r),
\end{equation}
where $\omega_{\infty}$  is the  light ray frequency measured by a static viewer at infinity,   while $\omega_{e}$ is the electron plasma frequency.
The  above refractive index simplifies to
\begin{equation}\label{25}
n(r)=\sqrt{1-\frac{\omega_{e}^{2}}{\omega_{\infty}^{2}}(\Lambda^\prime r^2-m_0-2q^2\ln\left(\frac{r}{l}\right)+m^2cc_1r)}.
 \end{equation}
For the given black hole, described by static
 spherically symmetric metric  surrounded by a plasma, the optical metric  is given by
\begin{equation}\label{26} 
d t^{2}=g_{i j}^{o p t} d x^{i} d x^{j}=n^{2}\left[\frac{d r^{2}}{f^{2}(r)}+\frac{r^{2} d \phi^{2}}{f(r)}\right].
 \end{equation}
 The   determinant of optical metric tensor  $g_{i j}^{o p t}$   is calculated as follows:
 \begin{eqnarray}
 \sqrt{\det (g^{opt}_{ij})}&=&\frac{1}{{\Lambda^\prime}^\frac{3}{2}r^2}(1-\frac{\omega_{e}^{2}}{\omega_{\infty}^{2}} \Lambda^\prime r^2)+\frac{m_0}{2{\Lambda^\prime}^\frac{5}{2}r^4}(3-\frac{\omega_{e}^{2}}{\omega_{\infty}^{2}}\Lambda^\prime r^2)\nonumber\\
 &+&\frac{q^2}{{\Lambda^\prime}^\frac{5}{2}r^4}\ln\left(\frac{r}{l}\right)(3-\frac{\omega_{e}^{2}}{\omega_{\infty}^{2}}\Lambda^\prime r^2)-\frac{m^2cc_1}{2{\Lambda^\prime}^\frac{5}{2}r^3}(3-\frac{\omega_{e}^{2}}{\omega_{\infty}^{2}}\Lambda^\prime r^2).
  \end{eqnarray}
  Gaussian curvature in the form of curvature tensor can be calculated as
 \begin{eqnarray} \label{28}
 \mathcal{K}&=&\frac{R_{r \phi r \phi}\left(g^{opt}_{ij}\right)}{\det \left(g^{o p t} _{ij}\right)},\nonumber\\
 & =&\frac{1}{\sqrt{\det \left(g^{opt}
 _{ij}\right)}} \left[\frac{\partial}{\partial \phi}\left(\frac{\sqrt{\det \left(g^{o p t} _{ij}\right)}}{ {g}_{r r}^{opt}}  {\Gamma}_{r r}^{\phi}\right)-\frac{\partial}{\partial r}\left(\frac{\sqrt{\det \left(g^{o p t}_{ij}\right)}}{ {g}^{opt}_{r r}}  {\Gamma}_{r \phi}^{\phi}\right)\right],\nonumber\\
 &=&\Lambda^\prime m_0(-1+\frac{\omega_{e}^{2}}{2\omega_{\infty}^{2}} \Lambda^\prime r^2-\frac{\omega_{e}^{4}}{\omega_{\infty}^{4}}\Lambda^\prime r^2+\frac{3\omega_{e}^{4}}{\omega_{\infty}^{4}}{\Lambda^\prime}^2 r^4 )+\Lambda^\prime q^2(3+\frac{\omega_{e}^{2}}{\omega_{\infty}^{2}}\Lambda^\prime r^2+ \frac{3\omega_{e}^{4}}{\omega_{\infty}^{4}}\Lambda^\prime r^2)\nonumber\\
&-&\frac{3\omega_{e}^{4}}{\omega_{\infty}^{4}}{\Lambda^\prime}^2 r^4+\Lambda^\prime q^2 \ln\left(\frac{r}{l}\right)(-2+\frac{2\omega_{e}^{2}}{\omega_{\infty}^{2}} \Lambda^\prime r^2+\frac{\omega_{e}^{2}}{\omega_{\infty}^{2}}r^2 -\frac{2\omega_{e}^{4}}{2\omega_{\infty}^{4}} \Lambda^\prime r^2)-\frac{q^2m_0}{r^2}
\nonumber\\
&+& \frac{6q^2m^2cc_1}{r}\ln\left(\frac{r}{l}\right)+\frac{2q^2m^2cc_1}{r}+O(q^4,m^4,c^2,c_1^2).
 \end{eqnarray}
 Now,  to calculate the bending angle in the weak field limit of the light ray using the
Gauss-Bonnet theorem  and to compare  it with non-plasma medium,
we consider it  at linear order only  as
 \begin{equation}\label{30}
 \tilde{\delta}=-\lim _{{\cal R} \rightarrow 0} \int_{0}^{\pi} \int_{\frac{b}{\sin \phi}}^{\cal R} \mathcal{K} d S.
 \end{equation}
Now, we calculate the quantity $\mathcal{K} d S$ as
 \begin{eqnarray}\label{31}
 \mathcal{K}dS&=&-\frac{(m_0-3q^2)}{{\Lambda^\prime}^\frac{1}{2}r^2}+\frac{m_0^2}{2{\Lambda^\prime}^\frac{1}{2}r^2}\frac{\omega_{e}^{2}}{\omega_{\infty}^{2}}-\frac{m_0}{{\Lambda^\prime}^\frac{3}{2}r^4}\left(\frac{3m_0}{2}-8q^2\right)\nonumber\\
 &-&\frac{3m_0^2}{2{\Lambda^\prime}^\frac{1}{2}r^2}\frac{\omega_{e}^{4}}{\omega_{\infty}^{4}}
 +\frac{m^2cc_1}{2{\Lambda^\prime}^\frac{3}{2}r^3}(3m_0-5q^2)+O(q^4,m^4,c^2,c_1^2).
 \end{eqnarray}
 With the help of  Eqs. (\ref{30}) and   (\ref{31}), we obtain the deflection angle of linearly charged massive BTZ black hole in plasma medium as
 \begin{eqnarray}
 \tilde{\delta}&=&\frac{2(m_0-3q^2)}{{\Lambda^\prime}^\frac{1}{2}b}-\frac{m_0^2}{2{\Lambda^\prime}^\frac{1}{2}b}\frac{\omega_{e}^{2}}{\omega_{\infty}^{2}}+\frac{4m_0}{9{\Lambda^\prime}^\frac{3}{2}b^3}\left(\frac{3m_0}{2}-8q^2\right)\nonumber\\
 &+&\frac{3m_0^2}{{\Lambda^\prime}^\frac{1}{2}b}\frac{\omega_{e}^{4}}{\omega_{\infty}^{4}}
 -\frac{m^2cc_1\pi}{8{\Lambda^\prime}^\frac{3}{2}b^2}(3m_0-5q^2).
 \end{eqnarray}
Here, we see that, similar to the previous case, the deflection angle calculated for the plasma medium also depends on parameters like $q, b, m$, and $\Lambda'$. 
 
 \section{Graphical analysis for plasma medium}\label{sec6}
 In this section, we analyse the behavior of deflection angle of charged massive black hole in plasma medium and its dependence on  various parameters.
  \subsection{Effect of impact parameter ($b$) on deflection angle ($ \tilde{\delta}$)}
\begin{figure*}[ht]
 \begin{center} 
 $\begin{array}{cccc}
\subfigure[]{\includegraphics[width=0.5\linewidth]{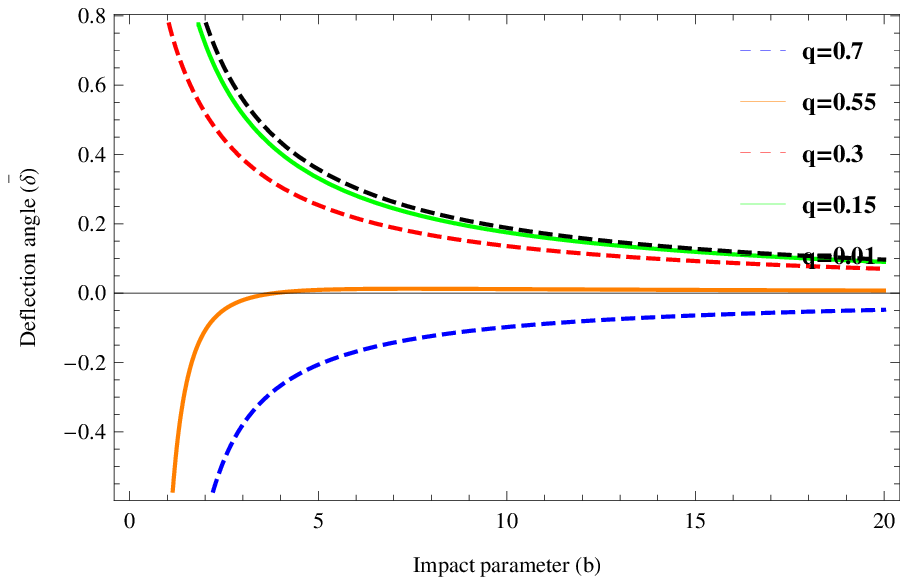}
\label{5a}}
\subfigure[]{\includegraphics[width=0.5\linewidth]{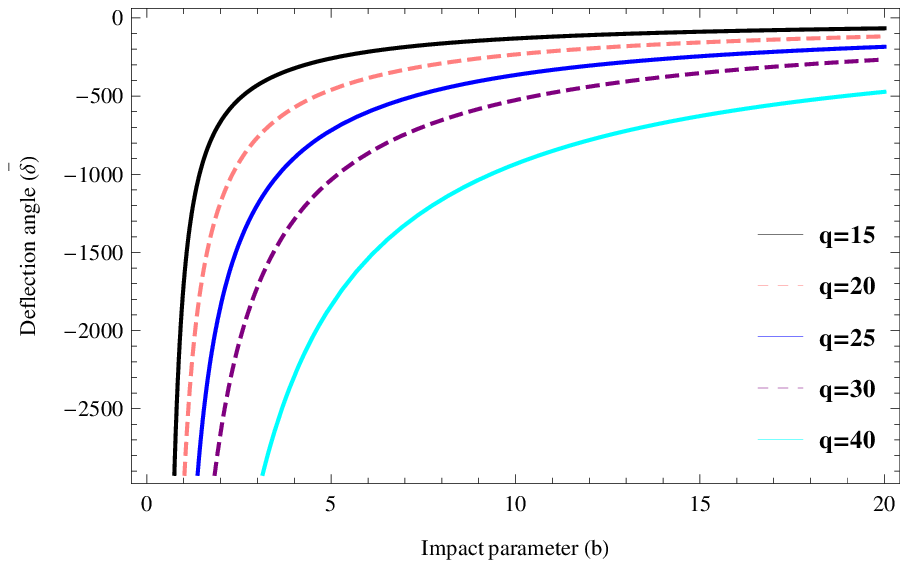}\label{5b}} \\
\subfigure[]{\includegraphics[width=0.5\linewidth]{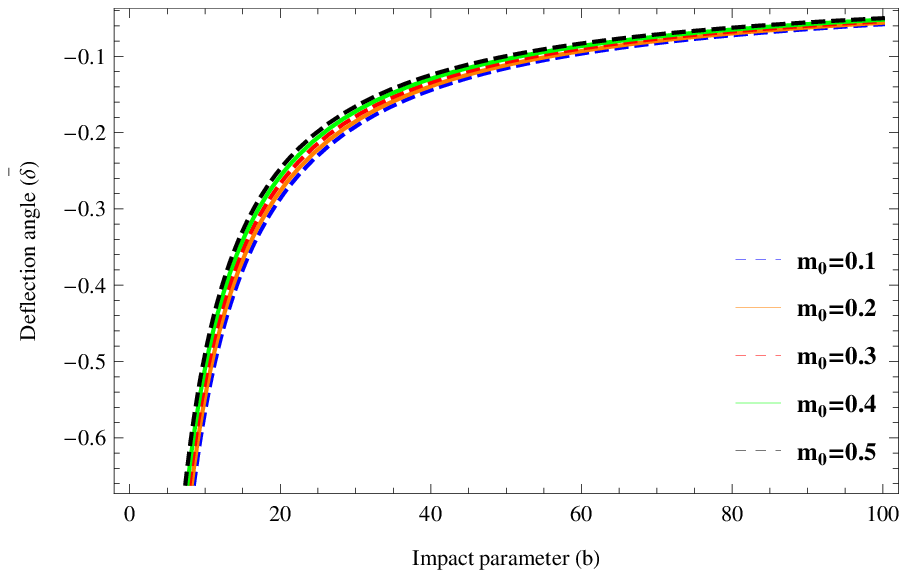}\label{5c}}
\subfigure[]{\includegraphics[width=0.5\linewidth]{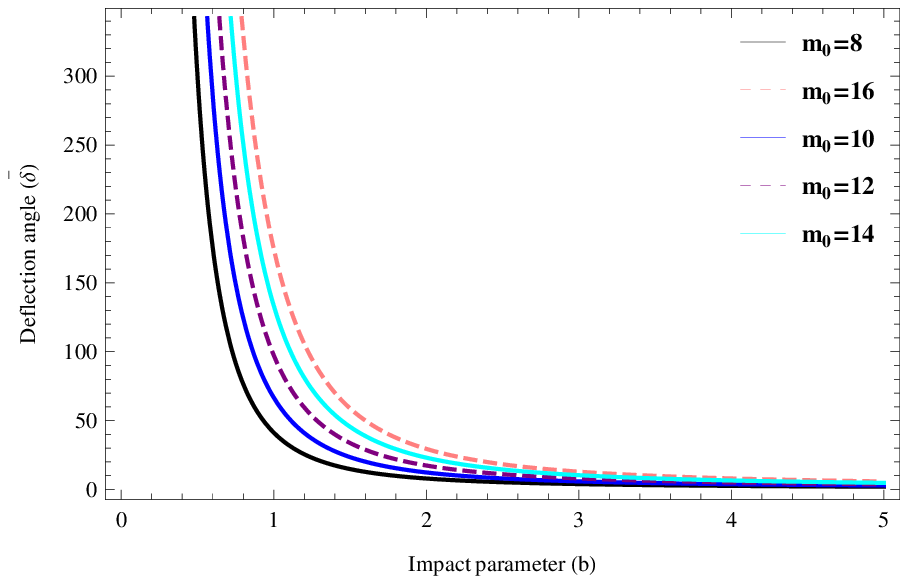} \label{5d}}
\end{array}$
\end{center}
\caption{   the behavior of deflection angle ($\tilde{\delta}$) with respect to impact parameter ($b$) for varying $q$ but fixed $m_0=1$ is depicted in  \ref{5a} and \ref{5b}. 
 The behavior of  $\tilde{\delta}$ with respect to $b$ for varying $m_0$ but fixed $q=1$ is depicted in  \ref{5c} and \ref{5d}. Here, $\Lambda'=c=c_1=1$. }
 \label{fig5}
\end{figure*}

  To study the behavior of deflection parameter and its dependence on impact parameter   for varying $q$ and $m_0$, we plot figure \ref{fig5}.
From the plots, we see that the deflection angle is an exponentially decreasing function of $b$ for large $m_0$ and very small $q$ but it takes always a positive value. However,
the deflection angle is an exponentially increasing function of $b$ for very small $m_0$ and large $q$ but it takes always negative values. For large values of  $b$, the deflection angle saturates.

\subsection{Effect of  charge ($q$) on deflection angle ($ \tilde{\delta}$)}
\begin{figure*}[ht]
\begin{center}
\includegraphics[width=0.5\linewidth]{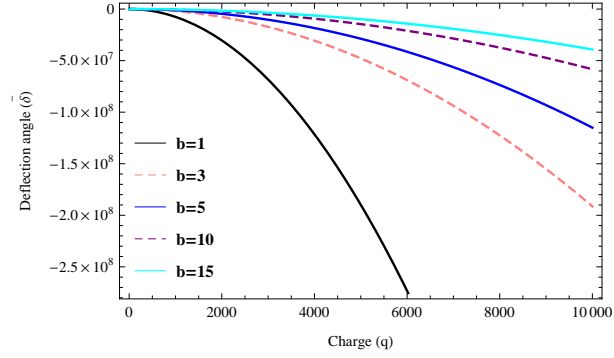}
\end{center}
\caption{The behavior of $\tilde{\delta}$ with respect to $q$  by changing impact parameter $b$.
Here, $b=m_0=\Lambda'=c=c_1=1$.}
\label{fig6}
\end{figure*}

From the plot \ref{fig6}, we see that the deflection angle is a decreasing function of $q$ in a plasma medium and becomes more negative for smaller $b$. 

 \subsection{Effect of mass parameter ($m$) on deflection angle ($ \tilde{\delta}$) }
\begin{figure}[ht]
   \begin{center} 
  $\begin{array}{cccc}
\subfigure[]{\includegraphics[width=0.5\linewidth]{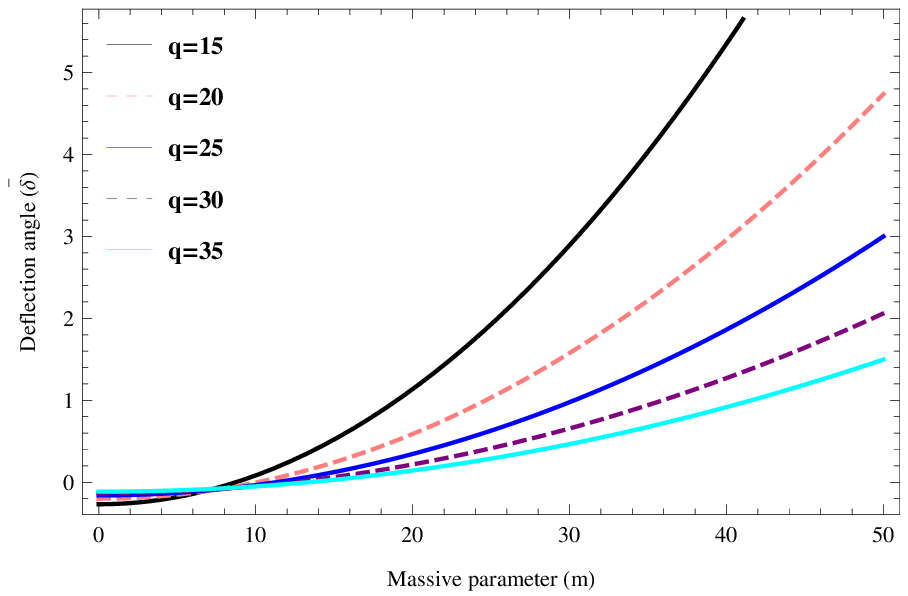}
\label{7a}}
\subfigure[]{\includegraphics[width=0.5\linewidth]{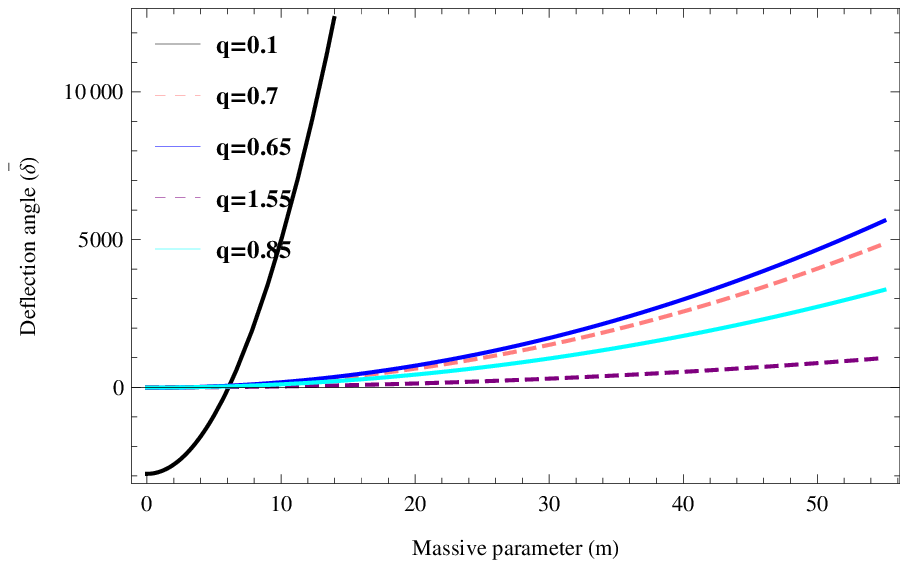}\label{7b}}  
\end{array}$
 \end{center}
\caption{The behavior of $\tilde{\delta}$ with respect to $m$  by changing charge $q$.
Here, $b=m_0=\Lambda'=c=c_1=1$.
 } 
\label{fig7}
\end{figure}

 The dependence of defection angle on mass parameter 
 by varying charge $q$ is depicted in figure \ref{fig7}.
 The deflection angle is an increasing function of mass parameter $m$.
In plot \ref{7a}, we see that, for smaller $q$ values, the deflection angle curve
decreases sharply. From figure \ref{7b}, for small $m$ and very small $q$, the deflection angle is negative valued. 

 \subsection{ Effect of cosmological constant ($\Lambda'$) on deflection angle ($ \tilde{\delta}$)} 
 \begin{figure}[ht]
 \begin{center}  $\begin{array}{cccc}
\subfigure[]{\includegraphics[width=0.5\linewidth]{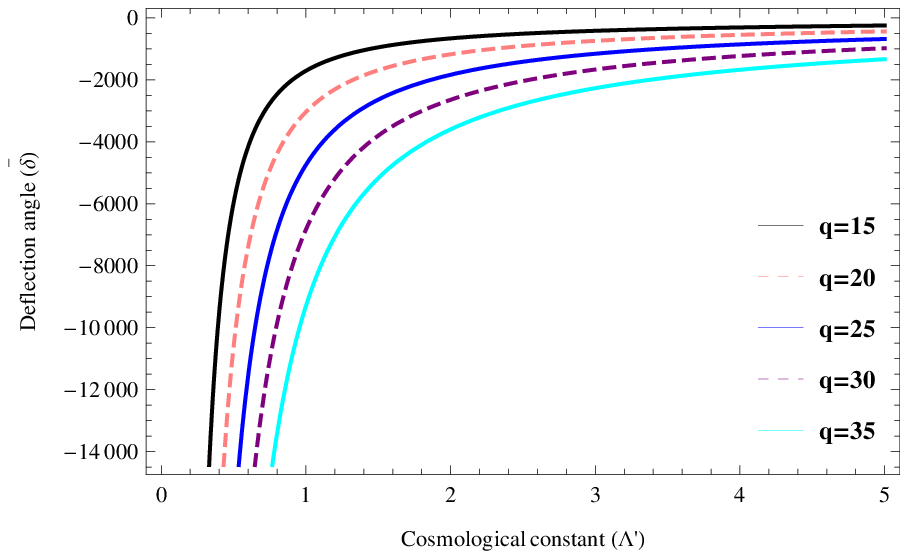}
\label{8a}}
\subfigure[]{\includegraphics[width=0.5\linewidth]{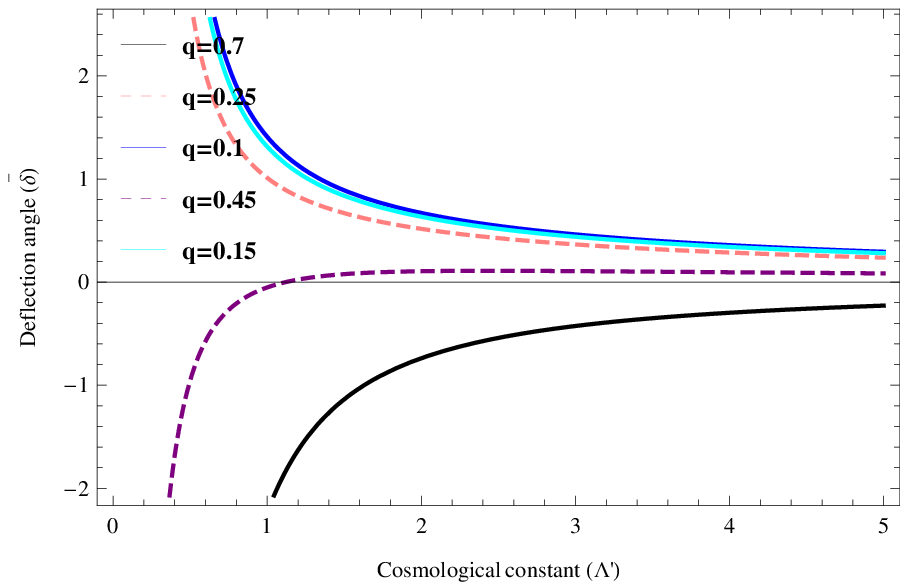}\label{8b}}  
\end{array}$
\end{center}
\caption{The behavior of $\tilde{\delta}$ with respect to $\Lambda'$  with changing $q$.  Here, $b=m_0=m=c=c_1=1$.  
 }
 \label{fig8}
\end{figure}
The behavior of deflection angle versus $\Lambda'$ is depicted in figure \ref{fig8}.
For large $q$, the deflection angle takes negative values but remains increasing function of $\Lambda'$.
However, for small $q$, the deflection angle is a positive but remains decreasing function of 
 $\Lambda'$.
 
 \section{Bound on greybody factor of linearly charged massive BTZ black hole}\label{sec7}
 Greybody factors (transmission probability) in black hole physics is a quantity related to the quantum nature of a black hole which 
 corrects the Planckian spectrum. The greybody factor describes the emissivity of the given black hole solution (non-perfect blackbody). Here, we determine the rigorous bound of the transmission probability of the linearly charged massive BTZ black hole.   
 The general bounds of  the greybody factor can be stated as \cite{48}
 \begin{equation}\label{34}
T \geq \operatorname{sech}^{2}\left(\frac{1}{2 \omega} \int_{-\infty}^{\infty} V(r) d r_{*}\right),
\end{equation}
where  $r_{*}$ denotes the  tortoise coordinate and $\omega$ is the quasinormal
mode frequency. 
 
  Event horizons (exterior and interior) can be calculated by   vanishing metric function  (i.e. $f(r)=0$). This gives 
 \begin{equation}\label{37}
 \Lambda^\prime r^2-m_0-2q^2\ln\left(\frac{r}{l}\right)+m^2cc_1r=0.
 \end{equation}
Considering $l=1$ and $\ln(r)=(r-1)$,   the solution of the above equation  can be written as
 \begin{equation}\label{38}
r_{\pm}=\frac{2q^2-m^2cc_1\pm \mathcal{G}}{2\Lambda^\prime},
 \end{equation}
 where $\mathcal{G}=\sqrt{(m^2cc_1-2q^2)^2-4\Lambda^\prime(2q^2-m_0)}$.

 Now, we analyze the Regge--Wheeler equation for angular momentum $l$ and derive rigorous bounds on the greybody factors. 
Let us define the Regge--Wheeler equation  as
\begin{equation}\label{41}
\left(\frac{d^{2}}{d r_{*}^{2}}+\omega^{2}-V(r)\right) \psi=0,
\end{equation}
where 
\begin{equation}
d r_{*}=\frac{1}{f(r)} d r ,\label{tor}
\end{equation}
and potential $V(r)$  in $3D$ is given as
\begin{equation}\label{42}
V(r)=-\frac{1}{4} \frac{f^{2}(r)}{r^{2}}+\frac{1}{2} \frac{f(r) f^{\prime}(r)}{r}+l^2 \frac{f(r)}{r^{2}}.
\end{equation}

In order to discuss bound value of transmission probability, we first write bound in the expression  (\ref{34}) with the help of (\ref{tor}) as
\begin{equation}\label{44}
T \geq \operatorname{sech}^{2}\left(\frac{1}{2 \omega} \int_{r_{+}}^{\infty} \frac{V(r) d r}{f(r)}\right).
\end{equation}
The lower bound  value of transmission probability thus leads to
\begin{equation}\label{45}
T\geq\operatorname{sech}^{2}\left[\frac{1}{2 \omega} \int_{r_{+}}^{\infty}\left(-\frac{1}{4} \frac{f(r)}{r^{2}}+\frac{(1)}{2} \frac{f^{\prime}(r)}{r}+\frac{l^2 }{r^{2}}\right)dr\right].
 \end{equation}
For the given metric function, mentioned in (\ref{37}), the above simplifies to 
\begin{eqnarray}
 T&\geq&\operatorname{sech}^{2}\left[\frac{1}{2 \omega}\left\{\frac{1}{4}(\Lambda^\prime r_{+}+\frac{m_{0}}{r_{+}}+\frac{2q^2}{r_{+}}(\ln(\frac{r_{+}}{l})+1)+m^2cc_1 \ln(r_{+})\right.\right.\nonumber\\&+&\left.\left.\frac{1}{2}(-2\Lambda^\prime r_{+}-\frac{2q^2}{r_{+}}-m^2cc_1 ln(r_{+}))+\frac{l^2}{r_{+}}\right\} \right].
 \end{eqnarray}
Plugging the value of $r_{+}$ from equation (\ref{38}),  the bound on the greybody factor changes  to 
 \begin{eqnarray}
 T&\geq & \operatorname{sech}^{2}\left [\frac{1}{2 \omega}\left\{\frac{3}{8}(m^2cc_1-2q^2-\mathcal{G})+\frac{m_{0}\Lambda^\prime}{2(2q^2-m^2cc_1+\mathcal{G})}\right.\right.\nonumber\\&+&\left.\left.\frac{\Lambda^{\prime}q^2}{ 2q^2-m^2cc_1+\mathcal{G} }\ln\left( \frac{2q^2-m^2cc_1+\mathcal{G}}{2\Lambda^{\prime}l}\right)  -\frac{1}{4}m^2cc_1 \ln \left(\frac{2q^2-m^2cc_1+\mathcal{G}}{2\Lambda^\prime}\right)\right.\right.\nonumber\\&+&\left.\left.\frac{\Lambda^{\prime}(2l^2-q^2)}{ 2q^2-m^2cc_1+\mathcal{G} }\right\} \right].
 \end{eqnarray}
{\bf Special Cases:}
 \begin{itemize}
 \item {\bf Case I:} If massive BTZ  black hole has no electric charge (i.e. $q=0$), then the bound  on the greybody factor  becomes  
 \begin{eqnarray}
 T&\geq & \operatorname{sech}^{2}\left[\frac{1}{2 \omega} \left\{\frac{3}{8}(m^2cc_1-\mathcal{G})+\frac{m_{0}\Lambda^\prime}{2(\mathcal{G}-m^2cc_1)}-\frac{1}{4}m^2cc_1 \ln  (\frac{\mathcal{G}-m^2cc_1}{2\Lambda^\prime} )\right.\right.\nonumber\\
 &+&\left.\left.\frac{2\Lambda^{\prime}l^2}{(\mathcal{G}-m^2cc_1)} \right\}  \right].
 \end{eqnarray}
 \item {\bf Case II:} In the massless limit ($m=0$), the bound  on the greybody factor 
 takes the following form:
 \begin{eqnarray}
 T&\geq &\operatorname{sech}^{2}\left [\frac{1}{2 \omega}\left\{-\frac{3}{8}(2q^2+\mathcal{G})+\frac{m_{0}\Lambda^\prime}{2(2q^2+\mathcal{G})}+\frac{\Lambda^{\prime}q^2}{(2q^2+\mathcal{G})}\ln\left(\frac{2q^2+\mathcal{G}}{2\Lambda^{\prime}l} \right)\right.\right.\nonumber\\&+&\left.\left. \frac{ \Lambda^{\prime}(2l^2-q^2)}{(2q^2+\mathcal{G})}\right\} \right].
 \end{eqnarray}
\end{itemize}

  \section{Comparative analysis of greybody factor}\label{sec8}
\begin{figure}[ht]
\begin{center}
 \includegraphics[width=0.5\linewidth]{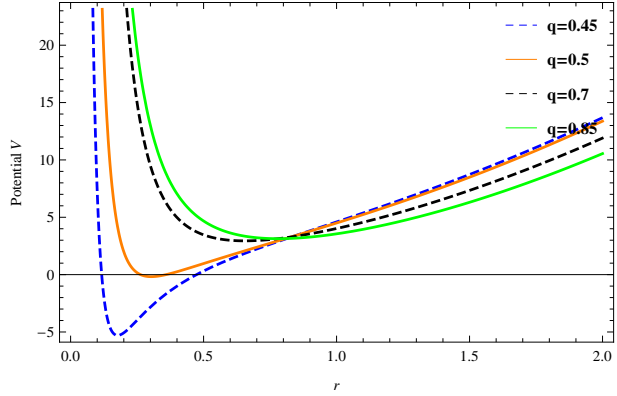}
\end{center}
\caption{ Potential ($V(r)$) versus $r$  for varying $q$  and for angular momentum ($l$) =1. Here, $\Lambda^{\prime}=m_{0}=m=c=c_{1}=l=1$.
 }\label{fig9}
 \end{figure}

  The behavior of potential is depicted in figure \ref{fig9} for different values of $q$. Here, we see that, for very small $q$, potential is negative which describes a bound system.
 
\begin{figure}[ht]
\begin{center}
\includegraphics[width=0.5\linewidth]{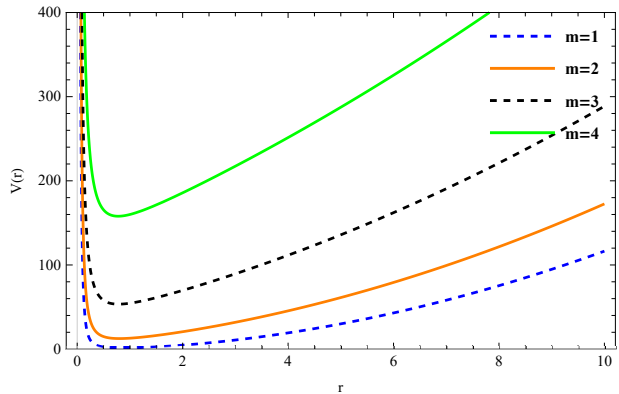}
\end{center}
\caption{Potential ($V(r)$) versus $r$  for varying $m$ and for angular momentum ($l$) =1. Here, $\Lambda^{\prime}=m_{0}=q=c=c_{1}=l=1$.}
\label{fig09}
 \end{figure}

   The behavior of potential is depicted in figure \ref{fig09} for different values of $m$. 

\begin{figure*}[ht]
   \begin{center} 
  $\begin{array}{cccc}
\subfigure[]{\includegraphics[width=0.5\linewidth]{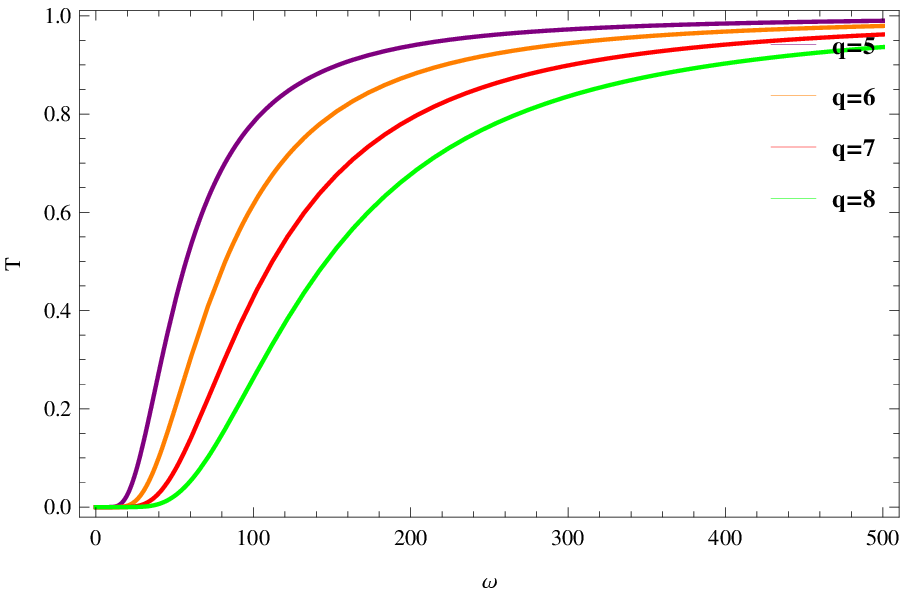}
\label{10a}}
\subfigure[]{\includegraphics[width=0.5\linewidth]{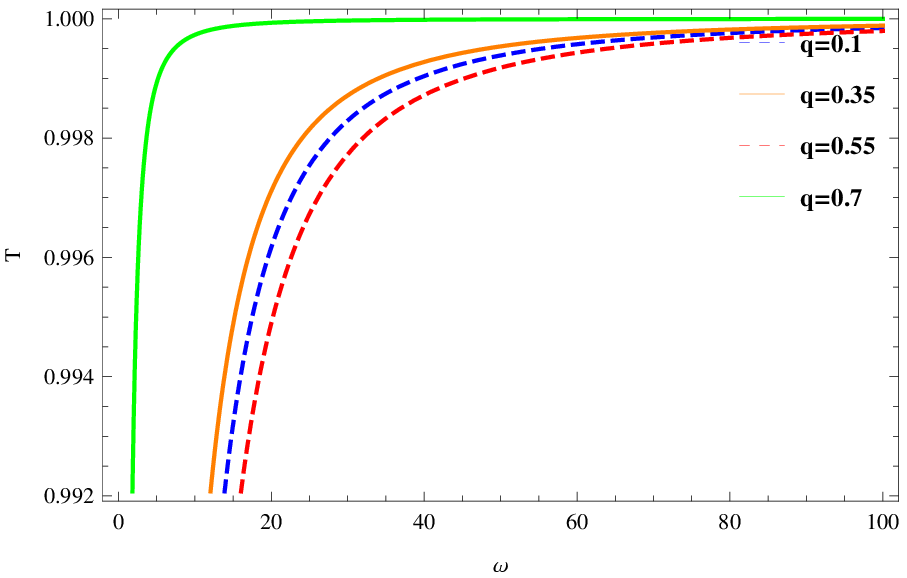}\label{10b}}  
\end{array}$
 \end{center}
 \caption{Plots for greybody factor lower bound with respect to $\omega$.}\label{fig10}
 \end{figure*}

 The calculated bound  with respect to $\omega$ is depicted in figure \ref{fig10}  for both small and large  values of $q$. We see that the bound increases sharply and then saturates after certain value of $\omega$. 
 \begin{figure*}[ht]
\begin{center} 
  $\begin{array}{cccc}
\includegraphics[width=0.5\linewidth]{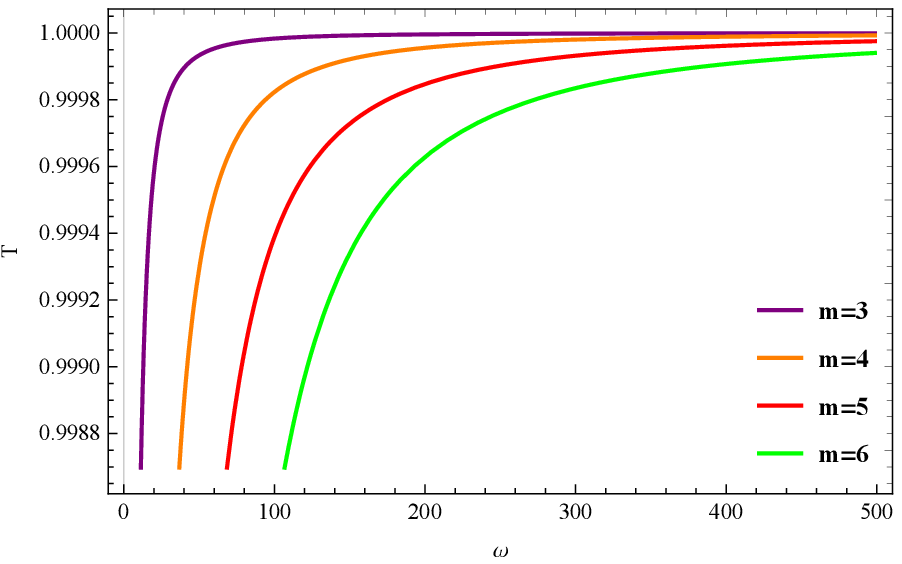} 
\end{array}$
 \end{center}
 \caption{Plots for greybody factor lower bound with respect to $\omega$ for changing $m$ but fixed angular momentum $l=1$. Here, $\Lambda^{\prime}= q=m_{0}=c=c_{1}=l=1$.}
 \label{fig010}
 \end{figure*}
 
 From figure \ref{10a}, we see that the value of bound decreases along with larger values of charge (in the large domain). In figure \ref{10b}, we observe that the value of bound increases along with larger values of charge (in the small domain).
 
\section{Shadow of charged massive BTZ black hole  }\label{sec09}
In order to study the null geodesics for linearly charged massive BTZ black hole, we use Hamilton-Jacobi method. To obtain the shadow of this black hole, we  calculate the celestial coordinates of the unstable null orbits. The motion of the particle on the linearly charged massive BTZ black hole is described by the following Lagrangian:
\begin{equation}\label{111}
\mathcal{L} =\frac{1}{2}g_{\mu\nu} u^\mu u^\nu.
\end{equation}
Here,   $u^{\mu}(= \frac{dx^{\mu}}{d\lambda})$ represents the four velocity of particle with affine parameter $\lambda$ along the geodesics. Since, the metric does not depend on the coordinate $t$ and $\phi$, so we get two constants of motion corresponding to these, namely, energy $E$ and angular momentum $h$  as follows
\begin{equation}\label{112}
E=p_{t}=\frac{d\mathcal{L}}{d\dot t}=g_{\phi t}\dot\phi+g_{tt}\dot t\; \ \ 
\mbox{and}\;\ \  h=-p_{\phi}=-\frac{d\mathcal{L}}{dt
\dot\phi}=-g_{\phi\phi}\dot\phi-g_{\phi t}\dot t.
\end{equation}
The derivatives of $t$ and $\phi$ with affine parameter $\lambda$ is derives as
\begin{equation}\label{113}
\frac{dt}{d\lambda}=\frac{E}{f(r)},\ \ \ \ \ \frac{d\phi}{d\lambda}=\frac{h}{r^2 \sin^2\theta}.
\end{equation}
The other geodesics equations can be estimated with the help of following relativistic Hamilton-Jacobi equation:
\begin{equation}\label{115}
\frac{dS}{d\lambda}= \frac{1}{2}g^{\mu\nu}\frac{dS}{dx^{\mu}}\frac{dS}{dx^{\nu}},
\end{equation}
where $S$ refers to  Jacobi action. In order to calculate the Hamilton-Jacobi equation, we consider an ansatz of the form \cite{sha}
\begin{equation}\label{116}
S=\frac{1}{2}m_{\star}^2\lambda-Et+h\phi+S_{r}(r)+S_{\theta}(\theta)
\end{equation}
where $m_{\star}$ is the mass of the test particle,   $S_{r}(r)$  and $S_{\theta}(\theta)$ correspond to the functions of $r$ and $\theta$.
In the charged massive  BTZ black hole spacetime, the Hamilton-Jacobi equation leads to
\begin{equation}\label{117}
\frac{1}{2}g^{tt}\frac{\partial{S}}{\partial{x^t}}\frac{\partial{S}}{\partial{x^t}}+g^{
\phi t}\frac{\partial{S}}{\partial{x^t}}\frac{\partial{S}}{\partial{x^{\phi}}}+\frac{1}{2}g^{rr}\frac{\partial{S}}{\partial{x^r}}\frac{\partial{S}}{\partial{x^r}}+\frac{1}{2}g^{\theta\theta}\frac{\partial{S}}{\partial{x^{\theta}}}\frac{\partial{S}}{\partial{x^{\theta}}}+\frac{1}{2}g^{\phi\phi}\frac{\partial{S}}{\partial{x^{\phi}}}\frac{\partial{S}}{\partial{x^{\phi}}}=-\dot S.
\end{equation}
Considering method of separation of variables, the solutions of $S_{r}$ and $S_{\theta}$ for null geodesic for massless particle  (photon) can be given, respectively,  as \cite{sha1}
\begin{eqnarray}
\Sigma \frac{\partial{S_{r}}}{\partial{r}}&=&\pm\sqrt{\mathcal{R}(r)},\label{118}\\
\Sigma \frac{\partial{S_{\theta}}}{\partial{\theta}}&=&\pm\sqrt{\Theta(\theta)}, \label{119}
\end{eqnarray} 
where 
$\mathcal{R}(r) = r^4E^2-(h^2+\mathbb{K})r^2f(r)$
and $\Theta(\theta) =\mathbb{K}-h^2\cot^2\theta $
with the  Carter  constant $\mathbb{K}$. For a far observer, the photon comes to charged massive BTZ black hole 
 near the equatorial plane and the unstable circular orbits follow: $\mathcal{R}(r)|_{r=r_{p}}=\mathcal{R}'(r)|_{r=r_{p}}=0$. Here, ``prime($'$)"
 denotes derivative with respect of $r$ and $r_p$ is  the radius of the unstable circular null orbit. 

Now, we introduce two dimensionless impact parameters $\xi$  and $\eta$ defined in terms of   $h$, $ E$ and $\mathbb{K}$, as
\begin{equation}\label{1112}
\xi=\frac{h}{E}  \ \ \mbox{and}\ \  \eta=\frac{\mathbb{K}}
{E^2}.
\end{equation}
In terms of these dimensionless impact parameters, the solution $\mathcal{R}(r)$ can be expressed as
\begin{equation}
\mathcal{R}(r) = r^4E^2-r^2E^2(\xi^2+\eta)f(r).\label{rr}
\end{equation}
Here, we note that $\mathcal{R}(r)$ acts as an effective potential for the photon  moving  along $r$ direction. 
With the help of relations (\ref{118}) and (\ref{rr}), the equation for $S_{r}$ can be written as
\begin{equation}\label{1113}
\left(\frac{\partial{S_{r}}}{\partial{r}}\right)^2+V_{eff}=0,
\end{equation}
where the effective potential $V_{eff}$ is given by
\begin{equation}\label{1114}
V_{eff}=\frac{1}{\Sigma^2} \left[r^2 {f(r)}{}E^2(\xi^2+\eta)-r^4E^2\right].
\end{equation}
Two conditions that unstable circular orbit follow, in terms of effective potential, turn to
\begin{equation}\label{1115}
\left. V_{eff}(r)\right|_{r=r_{p}}=0,\ \ \  \ \left.\frac{\partial{V_{eff}(r)}}{\partial{r}}\right|_{r=r_{p}}=0.
\end{equation}
For the give potential (\ref{1114}), the first  condition $V_{eff}(r)|_{r=r_{p}}=0$ gives 
\begin{equation}\label{1116}
\eta+\xi^2=\frac{r_{p}^2}{f(r_{p})}.
\end{equation}
The second boundary condition $\left.\frac{\partial{V_{eff}(r)}}{\partial{r}}\right|_{r=r_{p}}=0$, together with the first one, leads to
\begin{equation}\label{1117}
  r_{p}f^{\prime}(r_{p}) = 2f(r_{p}).
\end{equation}

 We can estimate the shadow size from Eq. (\ref{1116}) upon substituting the photon sphere radius
$r_{p}$. Using Eq. (\ref{1117}) and metric function of BTZ black hole, $r_{p}$ is calculated as 
\begin{equation}\label{1118}
r_{p}=\frac{6q^2-2m_{0}}{4q^2-m^2cc_{1}}.
\end{equation}
 The celestial coordinates  of the distant observer measured in
the directions perpendicular ($X$) and parallel ($Y$) to the projected
rotation axis describe the apparent angular distances of the image on the (celestial) sphere.  
For the present case, the celestial coordinates are given by
\begin{equation}\label{1119}
X= \lim_{r_{p}\to\infty} \left(\left.-r^2_{p}\sin\theta_{0}\frac{d\phi}{dr}\right|_{(r_p,\theta_0)}\right),
\end{equation}
and
\begin{equation}\label{1120}
Y= \lim_{r_{p}\to\infty} \left(\left. r^2_{p}\frac{d\theta}{dr}\right|_{(r_p,\theta_0)}\right),
\end{equation}
where $ \theta_{0}$ is the angular coordinate of the distant observer. For the null geodesic, this leads to 
\begin{equation}\label{1121}
X= -\frac{\xi}{\sin\theta},
\end{equation}
\begin{equation}\label{1122}
Y= \pm\sqrt{\eta-\xi^2 \cot^2\theta},
\end{equation}
which, in fact,  relate the
celestial coordinates  and impact parameters. 
In case when the observer is on the equatorial plane of the black hole    (i.e, $\theta_0=\frac{\pi}{2}$), the celestial coordinates take  the values:
\begin{equation}\label{1123}
X=-\xi,
\end{equation}
\begin{equation}\label{1124}
Y= \pm\sqrt{\eta}.
\end{equation}
Therefore,  the radius of the shadow can be given by
\begin{equation}\label{1125}
R_{s}=\sqrt{X^2+Y^2} =\sqrt{\eta+\xi^2}=\sqrt{\frac{r^2_{p}}{f(r_{p})}}.
\end{equation}
 Now, we demonstrate the computed values of the shadow radius $R_{s}$ for different values of cosmological constant ($\Lambda^{\prime}$) of the charged BTZ black hole in table \ref{table:1}. Table  \ref{table:2} shows the values of the shadow radius ($R_{s}$) for different values of charge ($q$).
\begin{table}[h!]
\parbox{.45\linewidth}{
\centering
\begin{tabular}{ |c| c| c| c| } 
 \hline
 $q$ & $r_{p}$ & $\Lambda^{\prime}$ & $R^2_{s}$ \\ [0.5ex] 
 \hline\hline
  &  & 0.45 & 3.80816 \\ 
  &  & 0.55 & 2.75791 \\
 1 & 1.333 & 0.67 & 2.07214 \\
  &  & 0.75 & 1.77748 \\
  &  & 0.84 & 1.4233 \\ [1ex] 
 \hline
  &  & 4.00 & 3.85718 \\ 
  &  & 4.15 & 2.44345 \\
 3 & 1.48571 & 4.25 & 1.9365 \\
  &  & 4.35 & 1.64134 \\
  &  & 4.45 & 1.40993 \\ [1ex] 
 \hline
\end{tabular}
\caption{Radius of the black hole shadow $R_{s}$ for varying $\Lambda^{\prime}$ with fixed $q$ and $r_{p}$.}
\label{table:1}}
\hfill
\parbox{.45\linewidth}{
\centering
\begin{tabular}{ |c| c| c| c| } 
 \hline
 $\Lambda^{\prime}$ & $q$ & $r_{p}$ & $R^2_{s}$ \\ [0.5ex] 
 \hline\hline
  & 1 & 1.333 & 0.262289 \\ 
  &1.5  & 1.4375 & 0.306845 \\
 4 & 2 & 1.4667 & 0.403015 \\
  & 2.5 & 1.47917 & 0.675016 \\
  &  2.8& 1.48353 & 1.29066 \\ [1ex] 
 \hline
   & 1 & 1.333 & 0.17204 \\ 
  &1.5  & 1.4375 & 0.190151 \\
 6 & 2 & 1.4667 & 0.223154 \\
  & 2.5 & 1.47917 & 0.287237 \\
  &  2.8& 1.48353 & 0.360386 \\ [1ex] 
 \hline
\end{tabular}
\caption{Radius of the black hole shadow $R_{s}$ for varying charge $q$ with fixed $\Lambda^{\prime}$. }
\label{table:2}}
\end{table}

\begin{figure}[h!]
\begin{center} 
 $\begin{array}{cccc}
\subfigure[]{\includegraphics[width=0.5\linewidth]{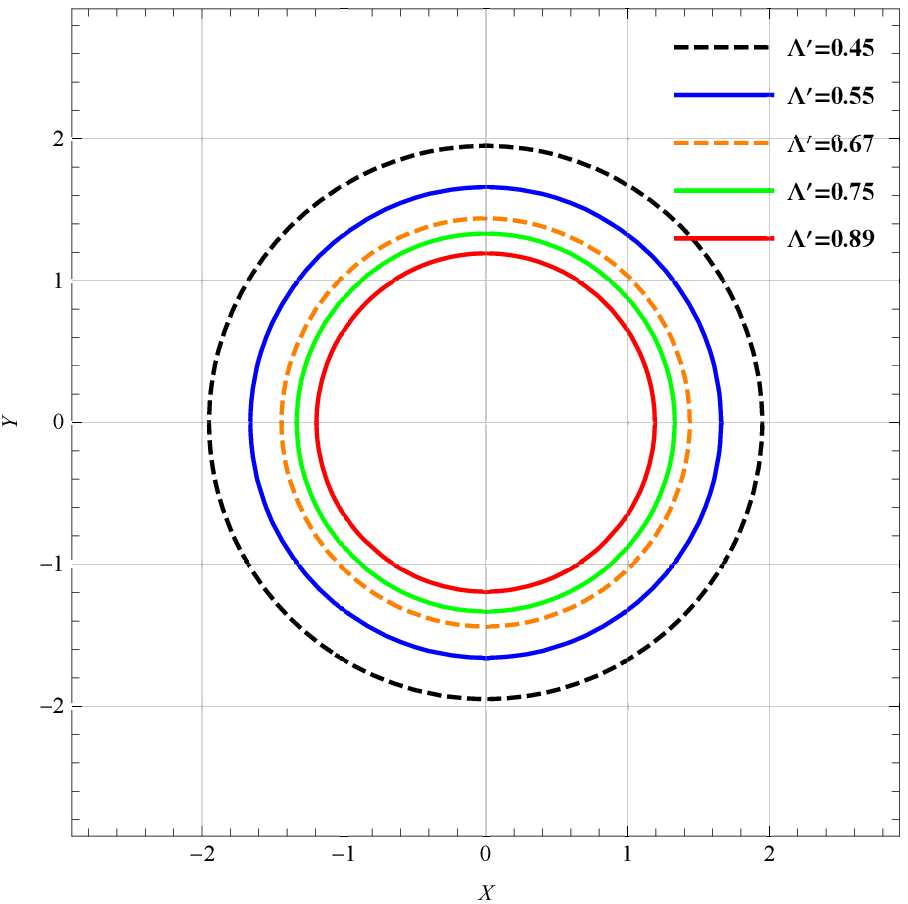}
\label{11a}}
\subfigure[]{\includegraphics[width=0.5\linewidth]{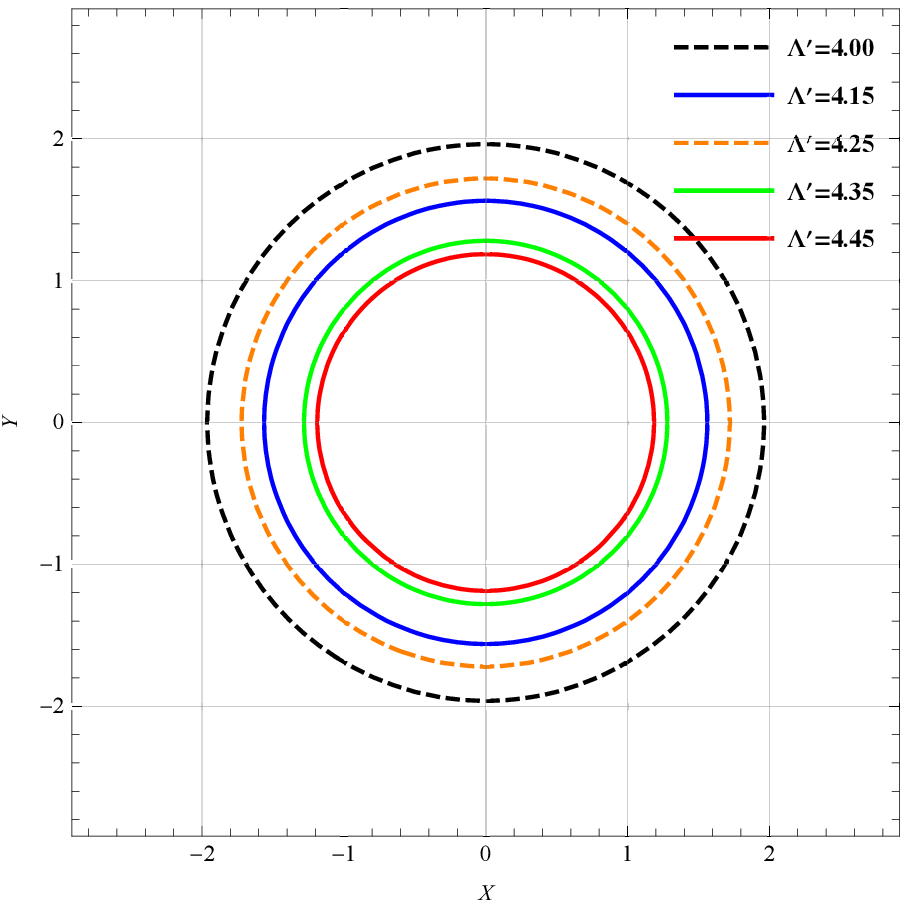}\label{11b}} 
\end{array}$
\end{center}
\caption{ Black hole shadow in the celestial plane   for different values of cosmological constant $\Lambda^{\prime}$. In (a),  charge $q=1$ and photon radius $r_{p} = 1.333$. In (b)  charge $q=3$ and photon radius $r_{p} = 1.48571$. Here $m_{0}=m=c=c_{1}=1$.  }
\label{fig11}
\end{figure}

\begin{figure}[h!]
\begin{center} 
 $\begin{array}{cccc}
\subfigure[]{\includegraphics[width=0.5\linewidth]{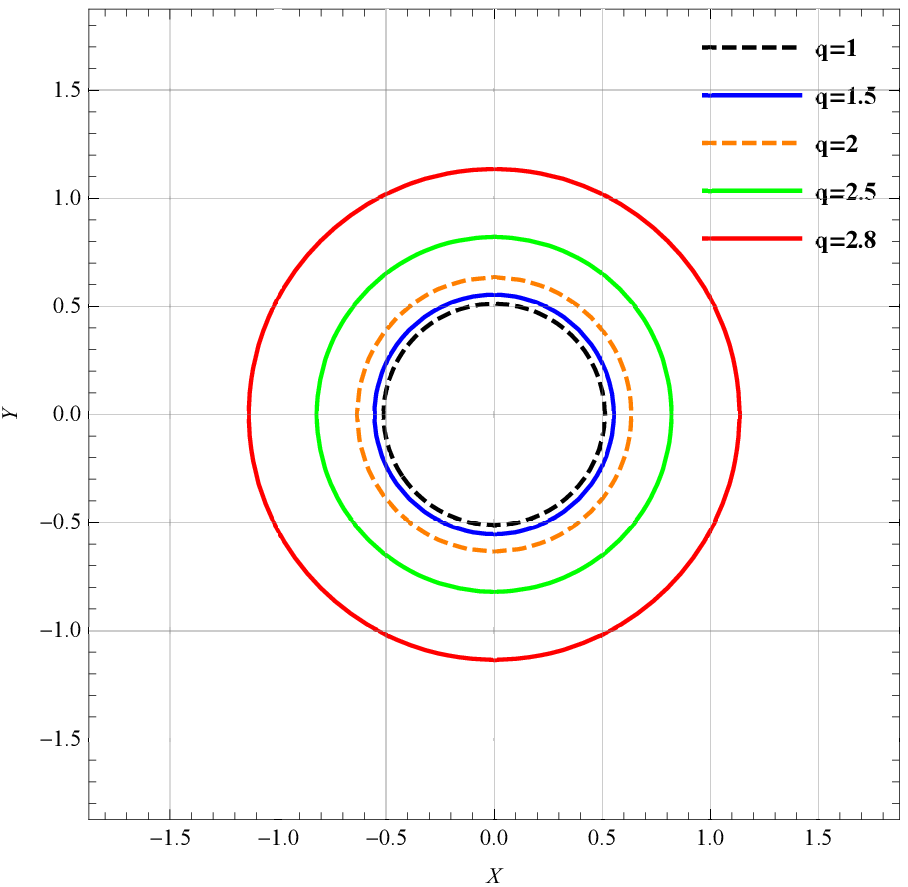}
\label{12a}}
\subfigure[]{\includegraphics[width=0.5\linewidth]{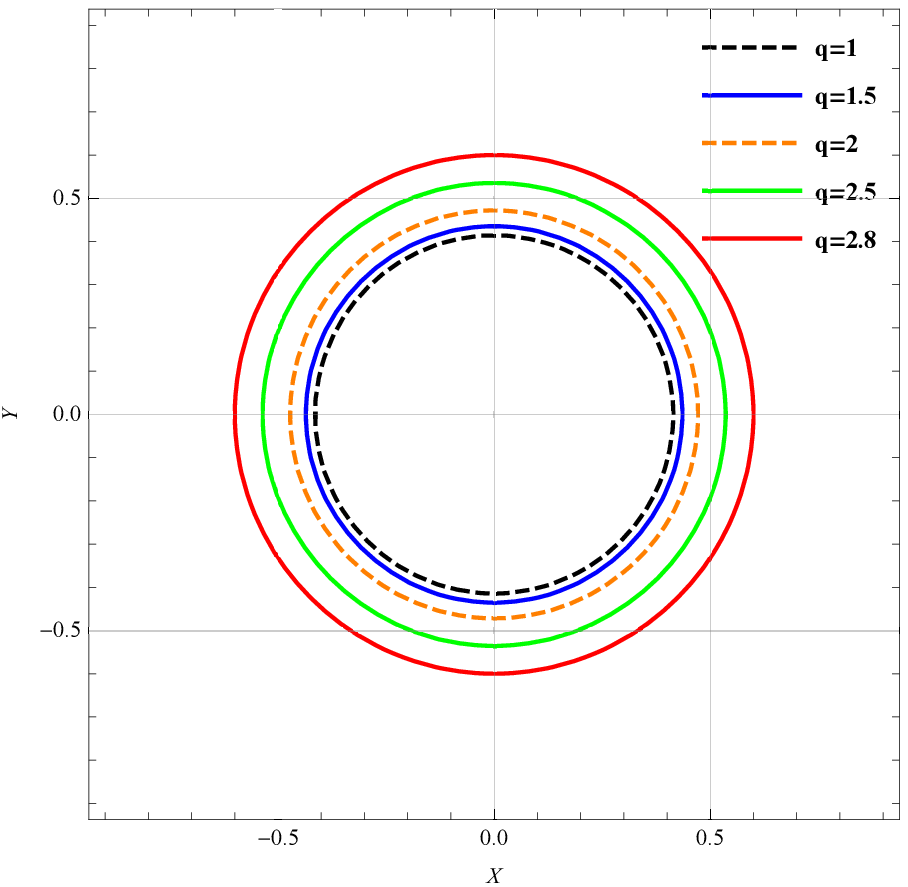}\label{12b}} 
\end{array}$
\end{center}
\caption{Black hole shadow in the celestial plane   for different values of charge $q$. In (a), $\Lambda^{\prime} = 4$.   In (b),  $\Lambda^{\prime} = 6$. Here $m_{0}=m=c=c_{1}=1$. }
\label{fig12}
\end{figure}

\begin{figure}[h!]
\begin{center} 
 $\begin{array}{cccc}
\subfigure[]{\includegraphics[width=0.5\linewidth]{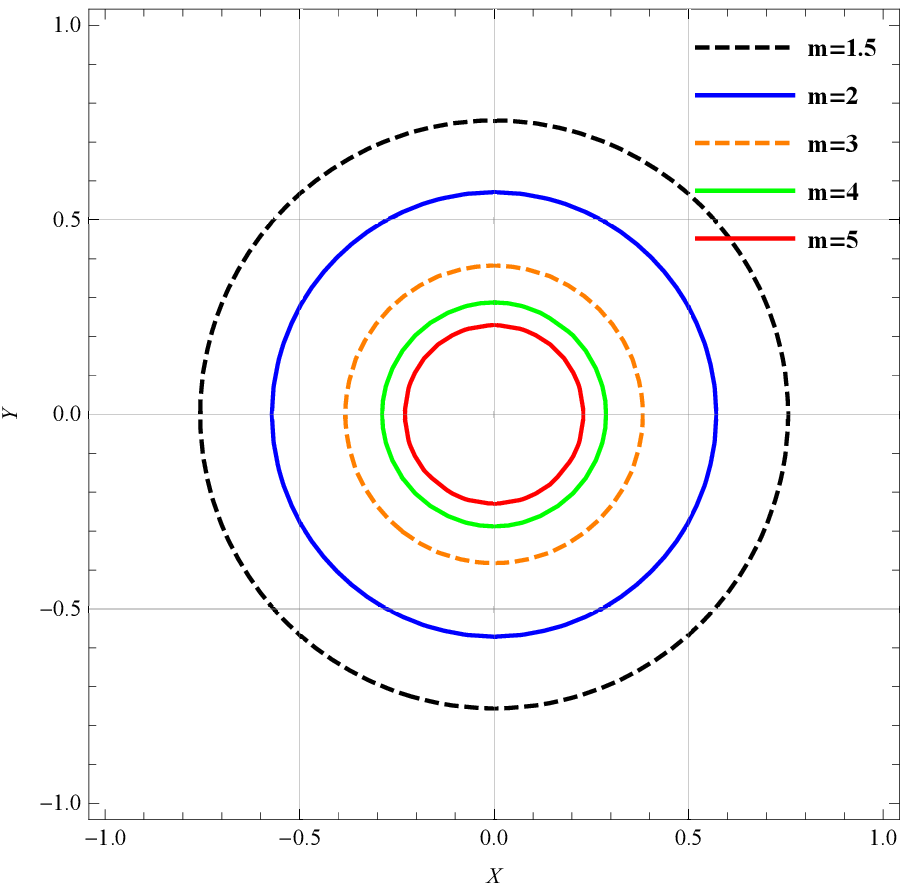}
\label{11a}}
\subfigure[]{\includegraphics[width=0.5\linewidth]{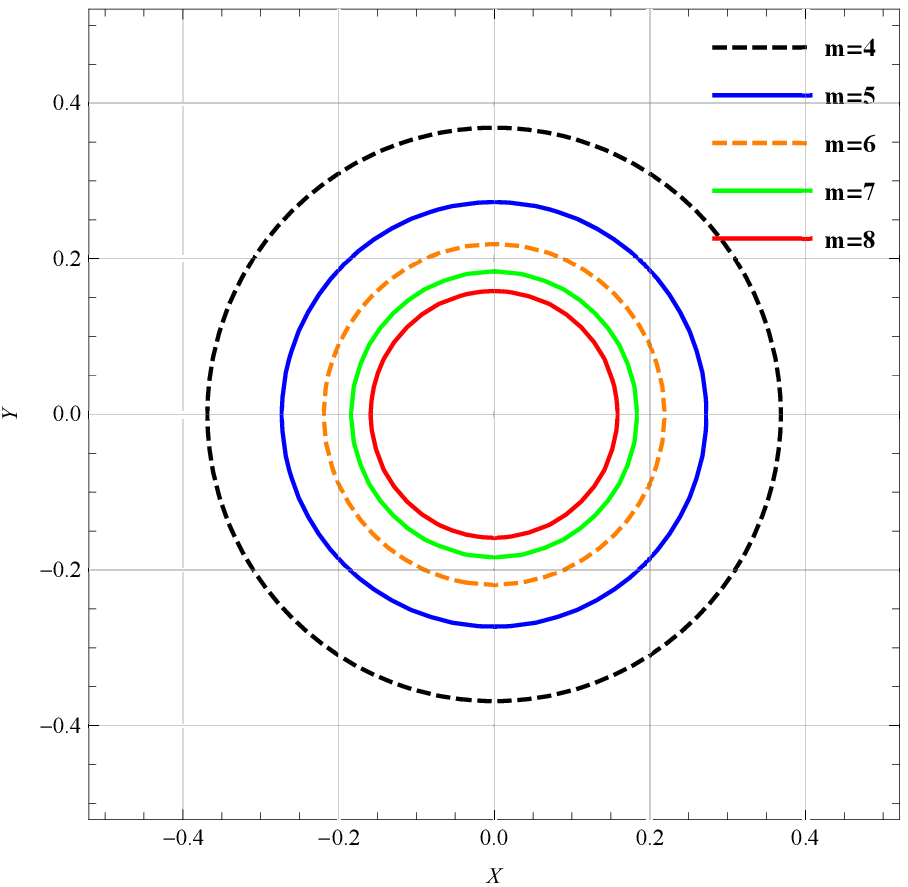}\label{11b}} 
\end{array}$
\end{center}
\caption{ Black hole shadow in the celestial plane   for different values of massive parameter $m$. In (a),  charge $q=1$ and photon radius $r_{p} = 1.333$. In (b)  charge $q=3$ and photon radius $r_{p} = 1.48571$. Here, $m_{0}=\Lambda^{\prime}=c=l=c_{1}=1$.  }
\label{fig011}
\end{figure}

\begin{figure}[h!]
\begin{center} 
 $\begin{array}{cccc}
\subfigure[]{\includegraphics[width=0.5\linewidth]{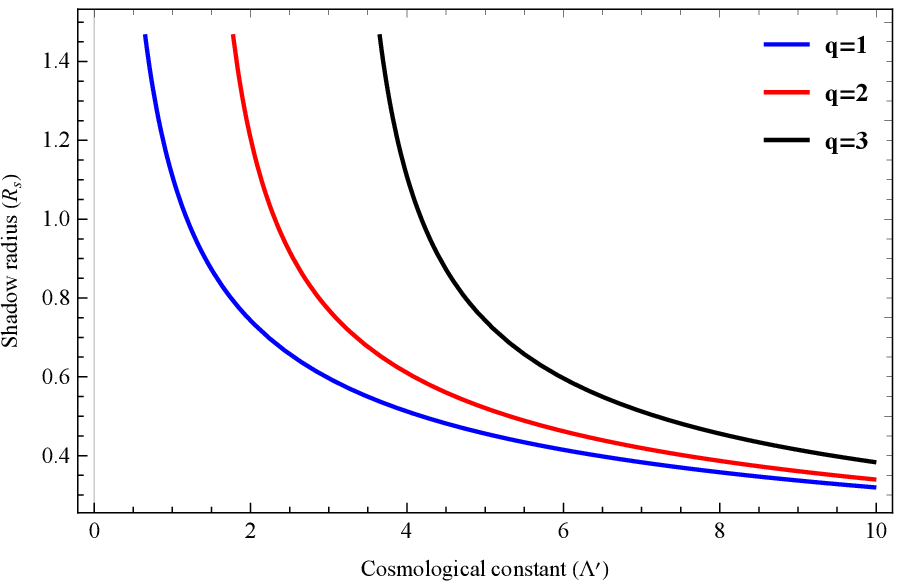}
\label{13a}}
\subfigure[]{\includegraphics[width=0.5\linewidth]{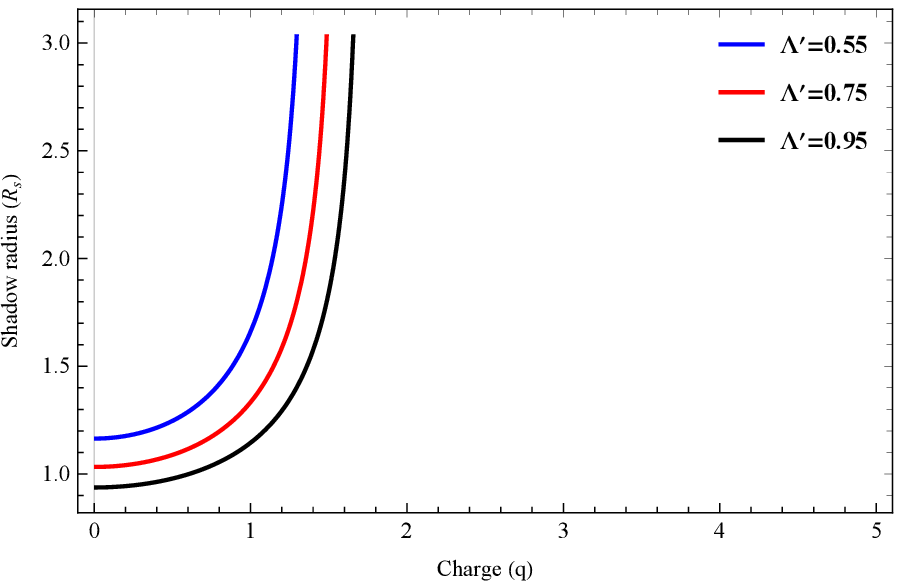}\label{13b}} 
\end{array}$
\end{center}
\caption{In (a): variation of the radius of the black hole shadow $R_{s}$ with cosmological constant $\Lambda^{\prime}$ for varying charge $q$. In (b): variation of the radius of the black hole shadow $R_{s}$ with charge $q$ for varying cosmological constant $\Lambda^{\prime}$.}
\label{fig13}
\end{figure}

\begin{figure}[h!]
\begin{center} 
 $\begin{array}{cccc}
\subfigure[]{\includegraphics[width=0.5\linewidth]{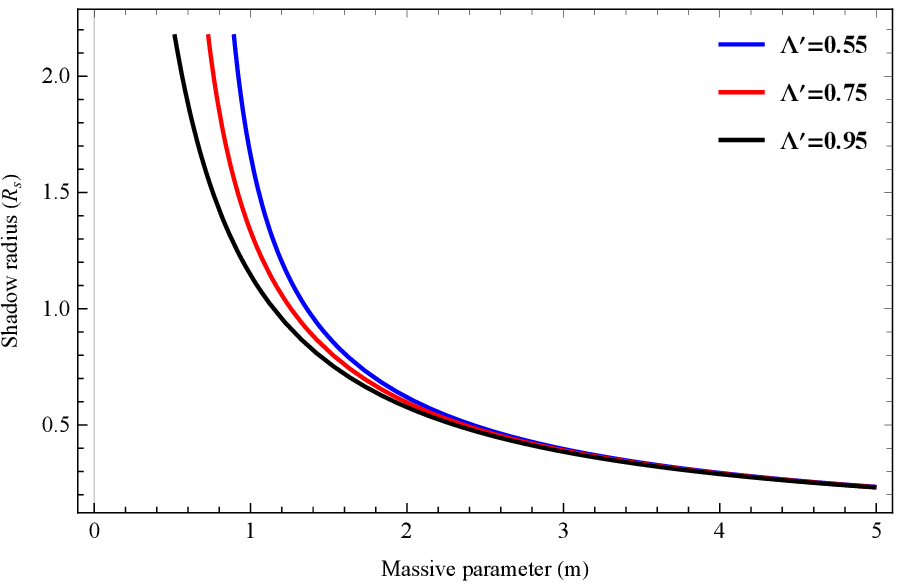}
\label{13a}}
\subfigure[]{\includegraphics[width=0.5\linewidth]{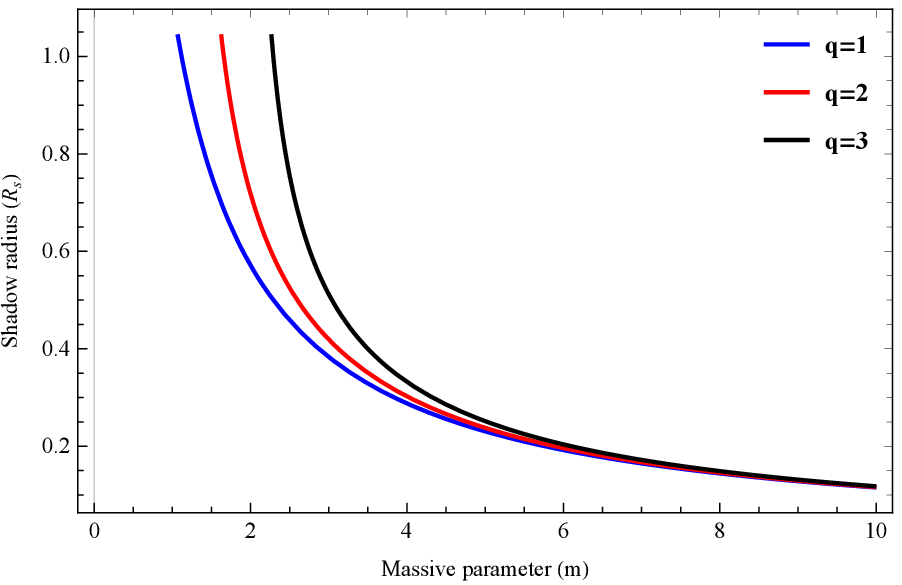}\label{13b}} 
\end{array}$
\end{center}
\caption{In (a): variation of the radius of the black hole shadow $R_{s}$ with massive parameter $m$ for varying cosmological constant $\Lambda^{\prime}$. Here, $q=m_{0}=c=l=c_{1}=1$. In (b): variation of the radius of the black hole shadow $R_{s}$ with massive parameter $m$ for varying charge $q$. Here, $\Lambda^{\prime}=m_{0}=c=l=c_{1}$. Both plots are for photon sphere radius $r_{p}=1.333$.}
\label{fig013}
\end{figure}

The black hole shadows for different values of cosmological constant, 
charge and mass parameter  are depicted in Figs. \ref{fig11}, \ref{fig12} and \ref{fig011}, respectively. 
Here, we observe that the shadow radius decreases with increase in cosmological constant and mass parameter. However, the shadow radius increases with increase in photon radius and charge. In figure \ref{fig13} and \ref{fig013}, we see that shadow radius is a decreasing function of cosmological constant and mass parameter, however, it is an increasing function of charge. 

 \section{Summary and final remarks}\label{sec9}
The existence of very compact objects (like black holes) is now well-established through astrophysical observations.
In the locality of black holes,
light ray travels through very strong gravitational fields.
 Black holes in a very unique sense provide an opportunity to study gravitational lensing beyond the first order weak deflection limit (which governs most gravitational lensing). However, the study of this new approach requires very strong technical abilities.

 In this work, we have considered a  charged black hole solution in massive gravity and with the help of null geodesics we have calculated optical metric. This optical metric 
 induces Gaussian optical curvature in weak gravitational lensing.
Applying the Gauss-Bonnet theorem,
  we  have calculated the deflection angle from the Gaussian curvature of the optical
metric for the black hole in a non-plasma medium. 
Here, we have found that the deflection angle of charged massive BTZ black hole depends on the various parameters like impact parameter,   charge, the mass parameter, graviton mass and cosmological constant.
To check the dependency of the deflection parameter on these parameters, we have done a graphical analysis. The graphs declared  that the deflection angle 
decreases and increases with the impact parameter for very small and large values of charge, receptively.  
However,   the deflection angle increases and decreases along with the impact parameter for small and large masses, respectively.   We also found that the deflection angle is a decreasing and increasing function of charge and mass of black hole, respectively.

  Within the context of gravitational lensing, we also computed Gaussian optical curvature for the charged massive BTZ black hole filled with
a cold spherically symmetric non-magnetized plasma.    The  Gauss-Bonnet theorem can be used to study the light rays in a plasma medium by following a correlation between timelike geodesics pursued by light rays in a plasma medium
and spatial geodesics in an associated optical geometry. The resulting
  deflection angle in the plasma medium has additional terms due to the refractive index of the plasma medium. We have provided the graphical analyses for the plasma medium case also, which reflects the dependencies of the deflection angle on various parameters. The rigorous analytic bounds on the greybody
factors of the linearly charged massive BTZ black hole are also derived.   From the potential and bound on greybody factor graphs, we observed that the system is not bound for a large value of charge and the bound increases sharply and then saturates after a certain value of quasinormal mode frequency.  Finally, we studied the shadow of charged massive BTZ black
holes for a   distant observer. The effects of charge and cosmological constant on the shadow radius are also analyzed. In this connection, we have found that  the shadow radius decreases with increase in cosmological constant. In contrast, the shadow radius increases with increase in photon radius and charge.

The present analysis may be useful  to probe the signature of  massive gravity  in the shadow of black holes which in turn may be a candidate for dark matter. The present analysis may be helpful in estimating correct value of  cosmological constant as a possible source of dark energy present in the universe. It will be interesting to  generalize these results
to the other gravity models such as  Lee-Wick gravity and $f(R)$ gravity. 
\section*{Acknowledgment}
This research was funded by the Science Committee of the Ministry of Science and Higher
Education of the Republic of Kazakhstan (Grant No. AP09058240).
 
 \section*{Data Availability Statement}
 
 Data sharing not applicable to this article as no datasets were generated or analysed during the current study.

\end{document}